\documentclass[sigconf]{acmart}

\usepackage{graphicx}
\graphicspath{{graphics/}}
\DeclareGraphicsExtensions{.pdf}
\usepackage[caption=false,font=footnotesize]{subfig}
\usepackage{algpseudocode} 
\usepackage{enumitem}
\usepackage[percent]{overpic}
\usepackage{balance}
\usepackage{tikz}

\copyrightyear{2017}
\acmYear{2017}
\setcopyright{acmlicensed}
\acmConference[DEBS '17]{ACM International Conference on Distributed and Event-Based Systems}{June 19 - 23, 2017}{Barcelona, Spain}
\acmPrice{15.00}\acmDOI{http://dx.doi.org/10.1145/3093742.3093914}
\acmISBN{978-1-4503-5065-5/17/06}

\begin{document}

\title[Minimizing Communication Overhead in Window-Based Parallel CEP]{Minimizing Communication Overhead in Window-Based Parallel Complex Event Processing}
\titlenote{Supported by Deutsche Forschungsgemeinschaft (DFG), project grant ``PRECEPT''.}


\author{Ruben Mayer, Muhammad Adnan Tariq, and Kurt Rothermel}
\email{{ruben.mayer, adnan.tariq, kurt.rothermel}@ipvs.uni-stuttgart.de}
\affiliation{%
  \department{Institute for Parallel and Distributed Systems}
  \institution{University of Stuttgart, Stuttgart, Germany}
}

\begin{abstract}

Distributed Complex Event Processing has emerged as a well-established paradigm to detect situations of interest from basic sensor streams, building an operator graph between sensors and applications. In order to detect event patterns that correspond to situations of interest, each operator correlates events on its incoming  streams according to a sliding window mechanism. 
To increase the throughput of an operator, different windows can be assigned to different operator instances---i.e., identical operator copies---which process them in parallel. This implies that events that are part of multiple overlapping windows are replicated to different operator instances. The communication overhead of replicating the events can be reduced by assigning overlapping windows to the same operator instance. However, this imposes a higher processing load on the single operator instance, possibly overloading it.
In this paper, we address the trade-off between processing load and communication overhead when assigning overlapping windows to a single operator instance.
Controlling the trade-off is challenging and cannot be solved with traditional reactive methods. 
To this end, we propose a model-based batch scheduling controller building on prediction. Evaluations show that our approach is able to significantly save bandwidth, while keeping a user-defined latency bound in the operator instances.

\end{abstract}



\maketitle

\begin{tikzpicture}
\begin{scope}[overlay]
\node[text width=40cm] at ([yshift=-19.5cm]current page.south) {(c) Owner 2017. This is the authors' version of the work. It is posted here for your personal use. Not for redistribution. \newline The definitive version is published in Proceedings of ACM International Conference on Distributed and Event-Based \newline Systems 2017 (DEBS '17), http://dx.doi.org/10.1145/3093742.3093914.};
\end{scope}
\end{tikzpicture}

\section{Introduction}
\label{sec:introduction}

Modern applications need to be able to react to situations occurring in the surrounding world. Thus, a growing number of sensor streams need to be processed in order to detect situations which the application or user is interested in, e.g., the traffic situation in a smart city or the detection of a person in a video surveillance application. 
To detect situations from sensor streams, Distributed Complex Event Processing (DCEP) \cite{Jain:2006:DIE:1142473.1142522, Schultz-Moller:2009:DCE:1619258.1619264} has been developed as a well-established paradigm building the bridge between sensors and consumers, i.e., applications or users that are interested in situations. A DCEP middleware deploys an operator graph in the network that incrementally detects patterns corresponding to situations in the sensor streams. In doing so, timeliness of pattern detection is of critical importance, as consumers need to react to occurring situations. This typically poses a soft latency bound on each operator of the DCEP system, because delayed situation detection leads to severe degradation of consumer benefits. For instance, late detection of a traffic jam leads to wrong routing decisions, and late detection of a person in a video surveillance application can mean that the relevant person has already left the scene.

In a DCEP system, high workload on the operators can lead to overload and long buffering delays when they process incoming streams only sequentially. To increase the operator throughput, data parallelization \cite{Balkesen:2013:RRI:2488222.2488257, 7024105} has been proposed as a powerful parallelization method. In a data parallelization framework, incoming event streams of an operator are split into windows that can be processed in parallel by an arbitrary number of operator instances, e.g., deployed in a cloud data center. To ensure consistency, each window comprises all events needed in order to detect a pattern. This means that different windows can overlap, i.e., events are part of multiple windows \cite{Balkesen:2013:RRI:2488222.2488257, 7024105}.

When splitting incoming event streams, the data parallelization framework assigns a window to an operator instance when the start of the window is detected. In doing so, assigning overlapping windows to different operator instances results in increasing communication overhead, as events that are part of multiple different windows are replicated to multiple operator instances. In the worst case, an event may be transmitted to all operator instances, leading to a high network load. In cloud data centers, this may not only impair the performance of the hosted DCEP systems, but also the performance of other applications hosted on the same infrastructure. Network-intensive applications have been identified as a major cause of bottlenecks in cloud data centers \cite{LaCurts:2013:CNT:2504730.2504744, Ballani:2011:TPD:2018436.2018465, Greenberg:2009:VSF:1592568.1592576}. Therefore, reducing the bandwidth consumption of parallel DCEP systems can be of great worth to all hosted applications.

To reduce the bandwidth consumption, we employ batch scheduling of subsequent overlapping windows, i.e., assigning them to the same operator instance. That way, events from the overlap only need to be transferred once. However, at the same time, the operator instance must process more windows in a shorter time. This can lead to temporary overload, so that events get buffered and queuing latency is accumulating. Nevertheless, latency between arrival of an event and its successful processing must not exceed a given latency bound. We address the following challenges in batching the optimal amount of windows, which cannot be solved with state-of-the-art scheduling algorithms from stream processing \cite{Carney:2003:OSD:1315451.1315523, Li:2005:NPN:1058150.1058158, balkesen2011scalable}.
\begin{itemize}[noitemsep,topsep=0pt]
	\item \textbf{Per-event latency}: Each incoming event at an operator can potentially trigger the detection of a pattern leading to a situation detection. Therefore, a latency bound should be kept for \emph{each single event}. 
	\item \textbf{Window overlap}: The overlap between windows of a batch influences the processing load induced by each event, as each event is processed in the context of each window it is part of. Moreover, the scheduling decision is made on open windows, i.e., the events and the overlap of a window are not known at scheduling time.
	\item \textbf{Automatic adaptation}: A batch scheduling controller should be able to automatically adapt to changing workload conditions without being manually trained for those conditions beforehand. 
\end{itemize} 

Toward this end, we make the following contributions in this paper. (1) Based on evaluations from different DCEP operators, we identify key factors that influence the latency in operator instances. In particular, we identify factors that have not been regarded in related work before. (2) Taking into account the identified key factors, we propose a model-based batch scheduling controller. The model allows to predict the latency induced in operator instances when assigning windows. (3) We provide extensive evaluations of the system behavior in two different scenarios, showing that our approach minimizes communication overhead while operator instances keep a required latency bound even when the system faces heavily fluctuating workloads.

\begin{figure}
\begin{minipage}[t]{0.35\linewidth}
    \includegraphics[width=\linewidth]{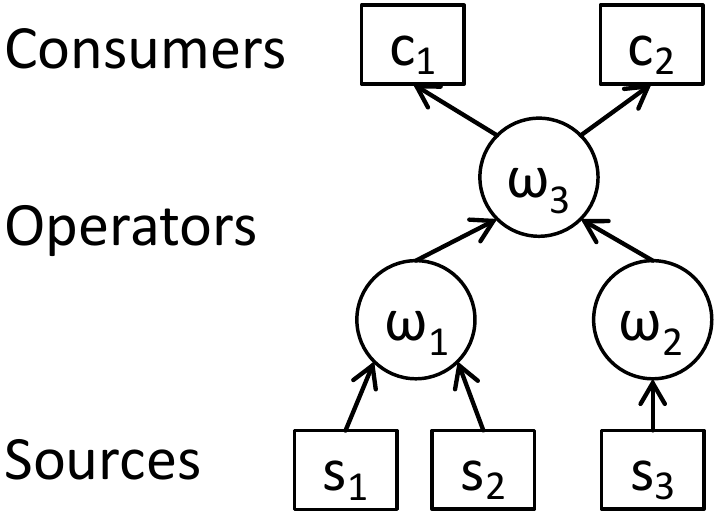}
\vspace{-0.3cm}
    \caption{DCEP operator graph.}
    \label{fig:cep_system}
\end{minipage}%
    \hfill%
\begin{minipage}[t]{0.6\linewidth}
    \includegraphics[width=\linewidth]{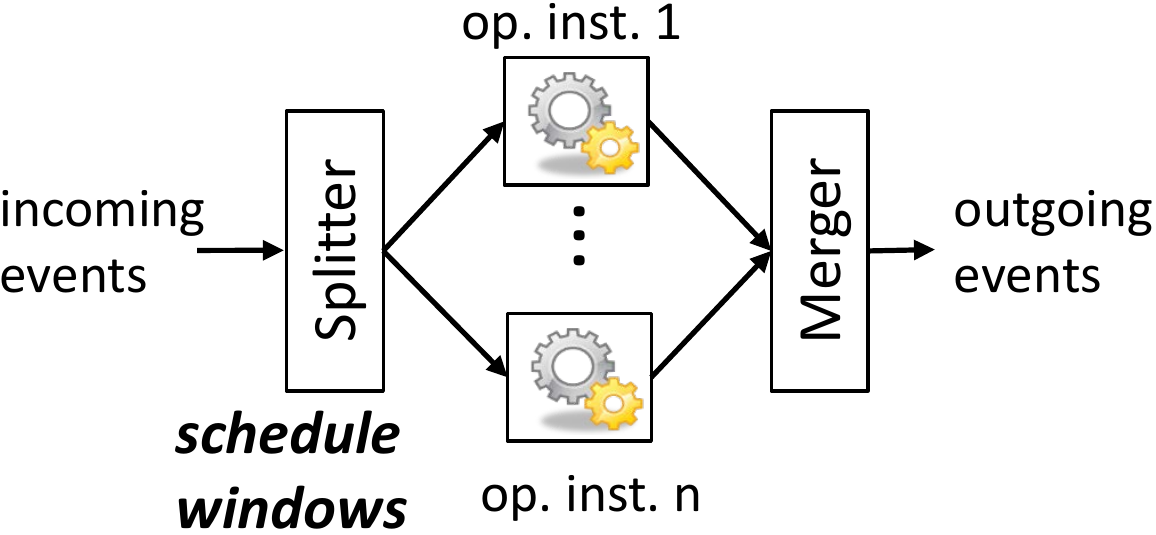}
\vspace{-0.3cm}
    \caption{Data parallelization framework.}
    \label{fig:parallel_execution_simplified}
\end{minipage}%
\vspace{-0.2cm}
\end{figure}

\section{Data-Parallel DCEP Systems}
\label{sec:System and Event Processing Model}

Before introducing the methods for bandwidth-efficient batch scheduling, we introduce a common model of a data-parallel DCEP system \cite{Balkesen:2013:RRI:2488222.2488257, 7024105}. 

A DCEP system builds an operator graph interconnecting event sources, operators and consumers by event streams. 
For example, Figure \ref{fig:cep_system} depicts a DCEP deployment with 3 sources, 3 operators and 2 consumers. 
An event $e$ consists of its payload and a header containing its event type. Events from all streams inherently have a well-defined total order\footnote{This order can, for instance, be established on time-stamps assigned by event sources with synchronized clocks, so that it reflects the ordering of physical occurrence of source events.}. When receiving events from different incoming streams, operators assign sequence numbers to the events according to the global order, process the events in-order and emit outgoing events to their successors in the operator graph. In doing so, an operator $\omega$ detects event patterns in finite, non-empty subsets of their incoming event streams---called \emph{windows} and denoted by $\textsf{w}$,
 using a correlation function $f_\omega : \textsf{w} \to (e_1, e_2, ..., e_n)$. The patterns to be detected can be defined in event specification languages like Snoop \cite{Chakravarthy1994} or Tesla \cite{Cugola:2010:TFD:1827418.1827427}, e.g., sequence patterns, logical (AND / OR / NOT) patterns, and others. Pattern definitions take into account event meta-data such as timestamps and event types, but can also rely on user-defined functions that analyze the events' payload, e.g., face recognition functions.

To cope with high workload, each operator is executed in a data parallelization framework (cf. Figure \ref{fig:parallel_execution_simplified}). It consists of a split--process--merge architecture \cite{Balkesen:2013:RRI:2488222.2488257, 7024105}. A splitter divides the incoming event streams of the operator into windows.  
The windows are then scheduled (i.e., assigned) to an elastic set of operator instances which simultaneously process their assigned windows. Finally, a merger orders the events emitted by the operator instances into a deterministic sequence.  Each window must comprise all events needed in order to detect a pattern instance.

\begin{figure}
\centering
\includegraphics[width=0.45\textwidth]{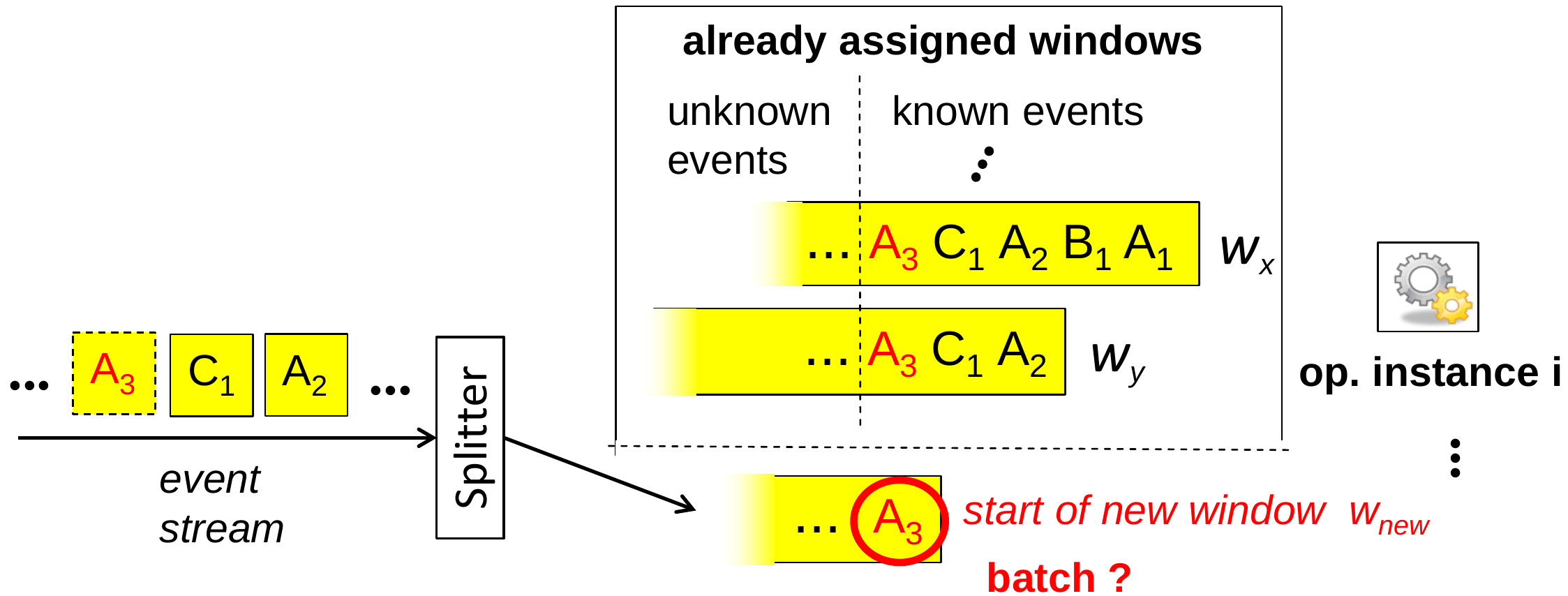}
\vspace{-0.2cm}
\caption{Splitting and scheduling.}
\label{fig:batching}
\vspace{-0.3cm}
\end{figure}

\textbf{Example:} In the scenario in Figure \ref{fig:batching}, the pattern to be detected is ``within one minute after occurrence of an event of type A, a sequence of events of type B and C occurs''---i.e., $\mathit{Aperiodic[A;}$\\$\mathit{Sequence(B;C);}$ $\mathit{A.timestamp}$ $\mathit{ + 1 min]}$ in Snoop syntax \cite{Chakravarthy1994}. The splitter opens a window whenever an event of type A occurs, and closes the window after one minute. The operator instances check whether in a window, events of type B and C occur in the right order. Taking a look at the splitting, we see that all events following A$_1$ within one minute are part of the same window $\textsf{w}_x$: If some of the events would be missing, they could not be checked for the Sequence(B;C) sub-pattern that follows A$_1$. 

In contrast to horizontal splitting, vertical splitting techniques (e.g., ``panes'' \cite{Li:2005:NPN:1058150.1058158} or ``stream batches'' \cite{Koliousis:2016:SWH:2882903.2882906}) have been proposed for stream aggregation operators. For instance, when the max or median value of a window of 1 minute shall be computed, that window could be split into 6 fragments of 10 seconds, the fragments' max or median be computed in parallel, and the global window's value be computed from the local results. This way, mere aggregation functions can be efficiently computed, as processing and aggregating the event subsets is an embarrassingly parallel task. In DCEP pattern detection, the processing of any event may depend on the complete temporal history of preceding events. For instance, when detecting a sequence of three events $A$, $B$, and $C$, processing of an event depends on the other events that have been detected before. Often, additional constraints are formulated, e.g., that $B$ and $C$ must occur within one minute from $A$ (as in the example above), or that $A$, $B$ and $C$ have a parameter $x$, such that $A.x < B.x < C.x$ (e.g., to detect chart patterns in stock markets \cite{Balkesen:2013:RRI:2488222.2488257}). This naturally leads to a window-based processing model, where windows capture the dependencies between different event sets that \emph{potentially} build a pattern match; the windows themselves can hardly be split into smaller fragments, because the dependencies between the events may span the complete window. Hence, horizontal splitting is a natural choice to exploit the data parallelism in such pattern detections (cf. \cite{Cugola:2010:TFD:1827418.1827427, 7024105}).

To allow for a virtually unlimited parallelization degree, all components are deployed on (possibly virtual) distinct shared-nothing hosts, and each of them can access a dedicated set of resources in terms of CPU and memory, i.e., we do not require shared memory between different operator instances or between the splitter and the operator instances.
The hosts of the components are inter-connected by unicast communication channels that guarantee eventual in-order delivery of streamed events. Focusing on the main technical challenges in this paper, we constrain ourselves to homogeneous hosts to deploy operator instances.

According to the pattern definition, windows can have different sizes and a different number of events can occur between two start events of subsequent windows. We denote the period of time that a window spans, i.e., the time between the first event and the last event of a window, as the \emph{window scope}, $\mathit{ws}$. Further, we denote the period of time between two start events of subsequent windows as the \emph{window shift}, $\Delta$. 
Upon detection of the start of a new window, this window is assigned to an operator instance according to a scheduling algorithm. In an operator instance, incoming events are processed sequentially. Within each window, an event has a different context. Therefore, when processing an event $e$, the operator instance sequentially processes $e$ in the context of each window that $e$ is part of. 

\textbf{Example:}  Recall the scenario in Figure \ref{fig:batching}. Two overlapping windows $\textsf{w}_x$ and $\textsf{w}_y$ have been assigned to the same operator instance $i$. When $i$ processes an event, e.g., $C_1$, this event has a different context in $\textsf{w}_x$ than in $\textsf{w}_y$: In $\textsf{w}_x$, the sequence (B;C) is detected, while in  $\textsf{w}_y$, the sequence is not detected. In checking the occurrences of the sequence pattern in different windows, operator instance $i$ processes $C_1$ sequentially first in $\textsf{w}_x$ and then in $\textsf{w}_y$.

From the event consumer's point of view, the situation detection latency is the period from the occurrence of a source event that signals a situation of interest until the situation is actually detected and signaled to the consumer. As the delayed detection of a situation degrades the benefits for the application, it poses a soft latency bound on the overall situation detection: violations of the latency bound shall, if possible, be avoided. 
The situation detection latency spans the whole operator graph and is sub-divided into latency budgets for each single operator. In each operator, the splitter and the merger induce latency for splitting the streams into windows and merging the results. Because scheduling windows to operator instances significantly influences the latency induced in each operator instance, in this paper, we focus on batch scheduling suitable amounts of windows to operator instances such that a latency bound in those operator instances is kept.

We define the \emph{operational latency} of $e$, $\lambda_o(e)$, as the period between the point in time when $e$ arrives at an operator instance and the point in time when $e$ is completely processed in all assigned windows in this operator instance. When, at the time of arrival of $e$, the operator instance is still busy with processing earlier events, $e$ waits in a queue until its processing can start. This is called \emph{queuing latency} of $e$, $\lambda_q(e)$. Then, $e$ is processed, which induces the \emph{processing latency} of $e$, $\lambda_p(e)$, the time from starting to process $e$ until $e$ is processed in all assigned windows. 
Overall, the operational latency of an event is a combination of its queuing latency and processing latency, i.e.,  $\lambda_o(e) = \lambda_q(e) + \lambda_p(e)$.

\textbf{Problem Formalization:}
To minimize the communication overhead, the batch scheduling controller tries to assign as many subsequent windows as possible to the same operator instance subject to the constraint that the operational latency of events in that instance must not exceed a latency bound $\mathit{LB}$. As soon as the start of a new window $\mathit{\textsf{w}_{new}}$ is detected by the splitter, the batch scheduling controller decides whether assigning that window to the same operator instance as the previous window would cause operational latency of events to exceed $\mathit{LB}$. This is noted as the \emph{batch scheduling problem} in data-parallel DCEP operators.

\textbf{Example:} The trade-off tackled in the batch scheduling problem is exemplified in Figure \ref{fig:batching}. An event $A_3$ arrives at the splitter and the splitter detects that $A_3$ starts a new window $\mathit{\textsf{w}_{new}}$ which now has to be scheduled. Suppose a set of previous windows $\mathit{\textsf{W}_{old}} = $(..., $\textsf{w}_x$, $\textsf{w}_y)$ has already been scheduled to a specific operator instance $i$. Events before $A_3$ in $\mathit{\textsf{W}_{old}}$ have been transferred to operator instance $i$. However, further events arriving after $A_3$ can as well be part of some of the windows in $\mathit{\textsf{W}_{old}}$; hence, they are transferred to operator instance $i$, too. When scheduling $\mathit{\textsf{w}_{new}}$ to operator instance $i$, communication overhead can be reduced, because events overlapping between $\mathit{\textsf{w}_{new}}$ and $\mathit{\textsf{W}_{old}}$ do not need to be transferred to multiple different operator instances. On the other hand, they need to be processed additionally in the scope of $\mathit{\textsf{w}_{new}}$, inducing higher processing latency. The splitter has to decide whether $\mathit{\textsf{w}_{new}}$ can be assigned to operator instance $i$ such that the operational latency does not increase beyond $\mathit{LB}$.

\begin{figure*}
  \centering
     \subfloat[Traffic monitoring: Processing latency of events in different positions in a window. $L1$ events: black, $L2$ events: red.]{
    \label{fig:poslat}
    \includegraphics[width=0.31\textwidth]{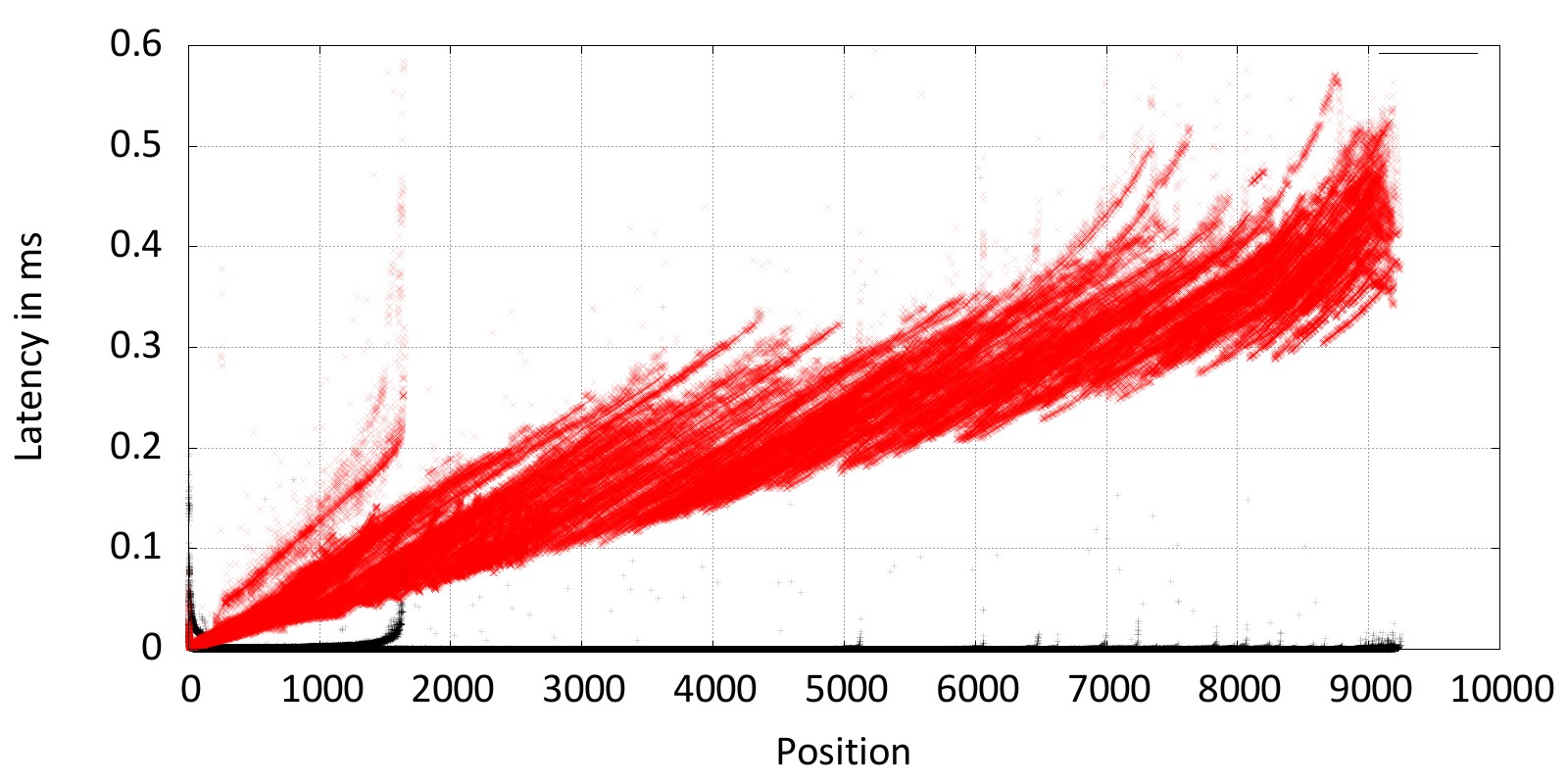} 
  }\ \  
 \subfloat[Face recognition: Processing latency of events in different positions in a window. Face events: black, query events: red.]{
    \label{fig:poslatvideo}
    \includegraphics[width=0.31\textwidth]{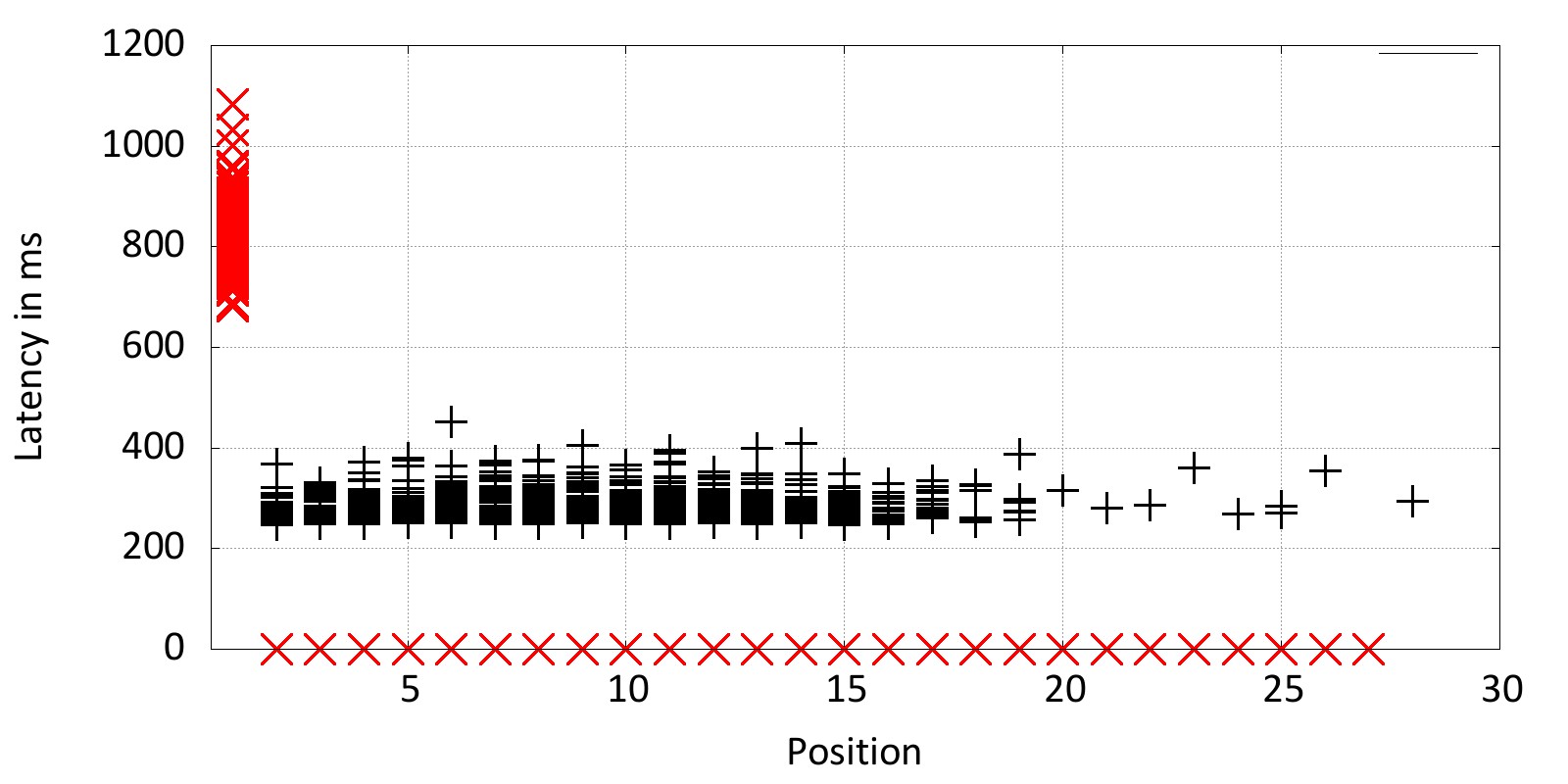}
  }\ \ 
  \subfloat[Traffic monitoring: Operational latency with reactive batch scheduling at $\mathit{TH} = 100 \mathit{ms}$ under different window scopes.]{
    \label{fig:eval/cumulated_latency_traffic_scenario}
    \includegraphics[width=0.26\textwidth]{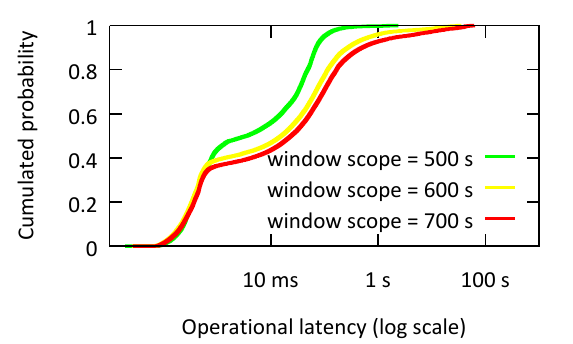}
  }
\vspace{-0.12in}
\caption{Evaluations.}
\vspace{-0.35cm}
\end{figure*}

\section{Batch Scheduling}
\label{sec:problemdescription}

To analyze the batch scheduling problem, in this section, we make the following contributions. First, in Section \ref{sec:controlproblem}, we identify and thoroughly analyze \emph{key factors} that influence the operational latency in an operator instance. 
We conclude that the impact of key factors on operational latency in an operator instance is complex and depends on the workload as well as on the operator.
Then, in Section \ref{sec:reactive}, we highlight the difficulties in developing a reactive batch scheduling controller that works without a latency model.

\subsection{Key Factors}
\label{sec:controlproblem}

In the following, we first identify and analyze key factors that influence the processing latency of events in the scope of a \emph{single} window. Based on that, we identify and analyze key factors that influence operational latency in a whole \emph{batch} of windows. To this end, we evaluate two different DCEP operators: a traffic monitoring and a face recognition operator. We ran all experiments on the computer cluster described in Section \ref{sec:evaluation} with a parallelization degree of 8.

\emph{Traffic monitoring operator}. A traffic monitoring application is interested in violations of an overtaking ban, so that the transgressor can be warned or punished. To this end, two cameras at two different locations ($L1$ and $L2$) on a highway capture video streams of vehicles passing by. To detect overtaking maneuvers, a traffic monitoring operator $\omega$ is deployed between the cameras and the application. When a vehicle passes a camera, an event is emitted to $\omega$, containing a time-stamp, the type (location $L1$ or $L2$), and the number plate. To detect the violations, $\omega$ uses an \emph{aperiodic} window window: Whenever a vehicle $a$ passes $L1$, a window $w$ is opened, and when the same vehicle passes $L2$, $w$ is closed. Another vehicle $b$ that appears in the $L1$ stream within $w$ has passed $L1$ after $a$. When $b$ appears again in $w$ in the $L2$ stream, it has passed $L2$ before $a$. If this is the case, $b$ has overtaken $a$ and thus violated the traffic rules. The query in $\omega$ can be expressed in CEP query languages, e.g., in Snoop \cite{Chakravarthy1994} language as an aperiodic operator: Aperiodic(A; B; C) with A $\to$ $\langle$plate=$a$, type=$L1$$\rangle$, B $\to$ Sequence($\langle$plate=$b$, type=$L1$$\rangle$;$\langle$plate=$b$, type=$L2$$\rangle$), C $\to$ $\langle$plate=$a$; type=$L2$$\rangle$\footnote{Aperiodic(A; B; C): Between the occurrence of two (complex) events A and C, the (complex) event B occurs.}.

\emph{Face recognition operator}. A face recognition application wants to know whether a person of interest is currently located in a specific area. To this end, pictures of detected faces from a camera are transferred to a face recognition operator $\omega$. Further, query events from users querying whether a certain person is in the current video stream are sent to $\omega$, containing a set of pictures of the person and a time frame within which the person shall be detected. $\omega$ uses a face recognition algorithm in order to detect whether the queried person is in the stream. This query can be resembled by an aperiodic operator Aperiodic(A; B; C) with  A $\to$ $\langle$type=\emph{query}, time=$t$$\rangle$, B $\to$ $\langle$type=\emph{face}, ``\texttt{face\_match(A)}''$\rangle$, C $\to$ $\mathit{time} \geq t + \mathit{time \ frame}$.

\textbf{Processing Latency of Events in a Window.}
\label{sec:Processing Latency of Events in a window}
When processing a \emph{single} window in an operator instance, each event imposes a specific processing latency. This is different from stream processing where the processing latency of an event in a window is considered fixed \cite{balkesen2011scalable, zeitler2011massive}. We identified two key factors that influence the processing latency of an event in a window: its type and its position. 

\emph{Event type}. Event types are a fundamental concept in DCEP. Many query languages, such as Snoop \cite{Chakravarthy1994}, Amit \cite{Adi:2004:ASM:988145.988150}, SASE \cite{wu2006high} and Tesla \cite{Cugola:2010:TFD:1827418.1827427}, allow for the definition of event patterns based on event types---e.g. SEQ(A;B), a sequence of events of type A and B. In the traffic monitoring operator, different event types are processed in a different way. $\mathit{L1}$ events are simply added to a list of seen events, while $\mathit{L2}$ events are compared to the seen events (cf. Figure \ref{fig:poslat}). In the face recognition operator, \emph{query events} are processed by building a face model of the queried person, while \emph{face events} are processed by comparing them to the established face model of the window (cf. Figure \ref{fig:poslatvideo}). In both operators, we see different processing latencies depending on the event types.

\emph{Position of Event}. When processing events of a window, internal state is gathered in an operator \cite{CastroFernandez:2013:ISO:2463676.2465282, Balkesen:2013:RRI:2488222.2488257}, which can influence the processing latency of events. For instance, in the traffic monitoring operator, an $\mathit{L2}$ event $e_{\mathit{L2}}$ can potentially complete a pattern Sequence($\langle$plate=$b$, type=$L1$$\rangle$;$\langle$plate=$b$, type=$L2$$\rangle$) or close the window. Therefore, $e_{\mathit{L2}}$ is compared to all $\mathit{L1}$ events that have been seen in the window before (\emph{equi-join} operator). Thus, with a higher position of  $e_{\mathit{L2}}$, its processing latency increases, as evaluated in Figure \ref{fig:poslat}. However, the processing latency of events does not necessarily increase with position. In the face recognition operator, each face event is compared to a query event; the \texttt{face\_match} function imposes the same processing latency in each event position (cf. Figure \ref{fig:poslatvideo}).

\begin{figure*}
  \centering
     \subfloat[Traffic monitoring operator.]{
    \label{table:1}
\footnotesize
\setlength{\tabcolsep}{0.3em}
\begin{tabular}{|l|l|l|l|l|l|l|l|}
\hline
\multicolumn{4}{|l|}{scenario parameters}                                                                                                       & \multicolumn{4}{l|}{measurements}                                                                                                                     \\ \hline
\begin{tabular}[c]{@{}l@{}}\# \end{tabular} & \begin{tabular}[c]{@{}l@{}}batch \\size \end{tabular} & \begin{tabular}[c]{@{}l@{}}avg. \\ iat (s)\end{tabular} &  \begin{tabular}[c]{@{}l@{}}$\mathit{ws}$ (s)\end{tabular} &\begin{tabular}[c]{@{}l@{}}max. op.\\ lat. (s)\end{tabular}  & \begin{tabular}[c]{@{}l@{}}feedback\\ delay (s)\end{tabular} &\begin{tabular}[c]{@{}l@{}}max. q.\\ length \end{tabular}&\begin{tabular}[c]{@{}l@{}}feedback\\ delay (s)\end{tabular} \\ \hline\hline  %
 1 &    500                     &                   0.15           &   900    &       2.4            &             725.5        &  15  &  773.6   \\ \hline    %
 2 &    500                     &                   0.125           &   900    &        3.7            &             757.7       &  27  &  724.8    \\ \hline    %
 3 &   500                     &                   0.1            &   900   &        24.1                 &             699.0    &  248  &  810.2       \\ \hline %
 4 &   750                  &                   0.1           &  900   &        100.6      &                      800.8      &  1029  &  844.0    \\ \hline  %
 5 &   1,000                  &                   0.1           &  900   &        116.1       &                     824.3       &  1194  &  795.6   \\ \hline  %
 6 &   1,000                  &                   0.1          &  1000  &          197.8         &                1,041.8         &  1699  &  999.0          \\ \hline %
 7 &   1,000                  &                   0.1          &  1100  &            199.2         &                1,179.2         &  1898  &  1,100.0         \\ \hline %
\end{tabular} 
  }\ \  
 \subfloat[Face recognition operator.]{
    \label{table:2}
\footnotesize
\setlength{\tabcolsep}{0.3em}
\begin{tabular}{|l|l|l|l|l|l|l|l|}
\hline
\multicolumn{4}{|l|}{scenario parameters}                                                                                                       & \multicolumn{4}{l|}{measurements}                                                                                                                     \\ \hline
\begin{tabular}[c]{@{}l@{}}\# \end{tabular} & \begin{tabular}[c]{@{}l@{}}batch \\size \end{tabular} & \begin{tabular}[c]{@{}l@{}}avg. \\ iat (s)\end{tabular} &  \begin{tabular}[c]{@{}l@{}}$\mathit{ws}$ (s)\end{tabular} &\begin{tabular}[c]{@{}l@{}}max. op.\\ lat. (s)\end{tabular}  & \begin{tabular}[c]{@{}l@{}}feedback\\ delay (s)\end{tabular} &\begin{tabular}[c]{@{}l@{}}max. q.\\ length \end{tabular}&\begin{tabular}[c]{@{}l@{}}feedback\\ delay (s)\end{tabular} \\ \hline\hline  %
 1 &    10                  &                   0.667         &   10    &    37.9         &     46.3     &  43  &  10.1     \\ \hline    %
 2 &    10                  &                   0.4       	&   10    &      68.7      &     77.1       &  84  &  9.4    \\ \hline    %
 3 &    10                  &                   0.286          &   10   &       99.7       &    108.0     &  115  &  10.6      \\ \hline %
 4 &    15                  &                   0.286          &   10   &        145.1      &        153.2    &  164  &  8.3    \\ \hline  %
 5 &    20                  &                   0.286          &   10   &        195.1       &       200.7     &  191  &  10.2     \\ \hline  %
 6 &    20                  &                   0.286          &   15  &        289.4         &     301.8      &  234 &  14.4       \\ \hline %
 7 &    20                  &                   0.286          &   20  &       392.1        &     410.1      &  258  &  19.6    \\ \hline %
\end{tabular}
 }
\vspace{-0.2cm}
\caption{Max. operational latency, queue length and feedback delays.}
\label{table:x}
\vspace{-0.3cm}
\end{figure*}

\textbf{Operational Latency in a Batch of Windows.}
In a \emph{batch} of windows, different windows may overlap. When the batch scheduling controller assigns a window to an operator instance that overlaps with other windows, the processing latency of all events in the overlap is influenced, as events are processed sequentially in the scope of their windows. Recall that a window has to comprise all events needed in order to detect a queried pattern. Therefore, the overlap of different windows cannot be changed by the batch scheduling controller. That is different from batch scheduling problems handled in stream processing, where batches are considered to be arbitrarily large, \emph{non-overlapping} sets of events, and batch scheduling decides how many \emph{events} shall be batched to a processing node \cite{Das:2014:ASP:2670979.2670995, Li:2005:NPN:1058150.1058158}. 

In the following, we identify key factors influencing the overlap of windows and analyze their impact on operational latency in operator instances. To this end, we run experiments with the traffic monitoring operator and the face recognition operator. In each experiment, using different traffic densities and different numbers of persons in a video frame, one key factor value is changed while all other key factors are kept constant, and the differences in operational latency peaks are analyzed (cf. Figure \ref{table:x}). For each experiment, more than 370,000 operational latency measurements have been taken. 

\emph{Batch size}. The batch size, i.e., number of windows assigned to an operator instance in a batch, influences the overlap of the windows, and hence, the operational latency of events. However, the relation between batch size and operational latency peak is not trivial. In the traffic monitoring operator, increasing the batch size by 50 \% and then by further 33 \% induces an increase in operational latency peak by 317 \% and 15 \%, respectively (cf. Figure \ref{table:1},  \#3, \#4 and \#5). In the face recognition operator, the relation between batch size and operational latency seems to be proportional (cf. Figure \ref{table:2}).

\emph{Inter-arrival time ($\mathit{iat}$)}. Given a fixed batch size, the inter-arrival time $\mathit{iat}$ of events influences the queuing latency of events. Further, it can influence the number of events in the windows, e.g., in time-based windows. The number of events in windows influences their overlap, which, in turn, influences the processing latency of the events. Thus, there is a complex relation between $\mathit{iat}$ and operational latency. In the traffic monitoring operator, we decreased the average $\mathit{iat}$ of events first by 17 \%, and then by further 20 \%. This induced an increase in operational latency peak by 54 \% and 551 \%, respectively (cf. Figure \ref{table:1}, \#1, \#2 and \#3). Similarly, in the face recognition operator, decreasing the average $\mathit{iat}$ of events first by 40 \% and then by further 28.5 \%, led to an increase in operational latency peak by 81 \% and 45 \%, respectively (cf. Figure \ref{table:2}).

\emph{Window scope ($\mathit{ws}$)}.
The window scope $\mathit{ws}$---i.e., the time between the start and end event of a window---depends on the queried patterns to be detected by the DCEP operator. It can be fixed to a specific time, e.g., when the query depends on a time-based window \cite{Arasu:2006:CCQ:1146461.1146463}, but it can also depend on the occurrence of specific events, e.g., in aperiodic queries or queries that define a sequence of specific events \cite{Cugola:2010:TFD:1827418.1827427, Chakravarthy1994}.  For instance, in the traffic monitoring operator, the start and end of a window depend on the speed of the vehicles, as a window starts when a vehicle passes $\mathit{L1}$ and ends when the same vehicle passes $\mathit{L2}$. When the speed of a vehicle is lower, the time spanned by the window opened from this vehicle is larger. 
Therefore, the size and overlap of windows can change even when the batch size and $\mathit{iat}$ stay the same. This is different from stream processing, where only windows of fixed size and fixed slide--- time- or count-based---are analyzed \cite{balkesen2011scalable}. 
In the traffic monitoring operator, we increased $\mathit{ws}$ in the traffic monitoring operator by 11 \%, and then by further 10 \%. This induced an increase in operational latency peak by 70 \% and 1 \%, respectively (Figure \ref{table:1}, \#5, \#6 and \#7). In the face recognition operator, however, increasing $\mathit{ws}$ led to a proportional increase in operational latency peaks.

From the observations on key factors that influence operational latency when processing a batch of windows, we see that building a direct mapping from \emph{batch size}, \emph{inter-arrival time} and \emph{window scope} to operational latency peaks in operator instances is hard. The relation between key factors and operational latency peaks that occur in operator instances is complex, and different in different operators. A model trained before run-time (off-line), hence, does not suffice; due to the complex relations between key factors, it is hard to train a model that can make reliable predictions outside of the learned parameter value ranges. Further, domain knowledge alone is not enough in order to hand-craft a latency model: Knowledge about the operator implementation does not necessarily help in understanding the relations between the identified key factors and the operational latency peak. 

In the following, we discuss whether the need for a latency model predicting the operational latency can be completely avoided by employing a reactive batch scheduling controller.

\subsection{Reactive Controllers}
\label{sec:reactive}

Here, we discuss the difficulties involved in devising a reactive batch scheduling controller. Reactive controllers are widely used in scheduling algorithms in the related field of parallel stream processing systems \cite{Das:2014:ASP:2670979.2670995, Li:2005:NPN:1058150.1058158}. The basic idea of a reactive controller is that it schedules windows according to \emph{feedback parameters} (like operational latency or queue length) from the operator instances that indicate how many windows can be batched. In the following, we point out the differences in batch scheduling in data-parallel DCEP operators to scheduling problems that have been solved with reactive controllers.
Then, we analyze operational latency and queue length of operator instances in the scope of the scenarios described in Section \ref{sec:controlproblem} in detail and show that none of these parameters provides reliable feedback to implement a reactive controller.

\begin{figure*}
  \centering
     \subfloat[Latency gains.]{
    \label{fig:latency_gains}
    \includegraphics[width=0.15\textwidth]{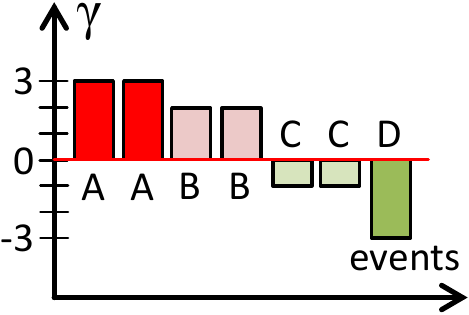} 
  }\ \  
 \subfloat[W-C: Latency peak.]{
    \label{fig:latency_queue_wc}
    \includegraphics[width=0.19\textwidth]{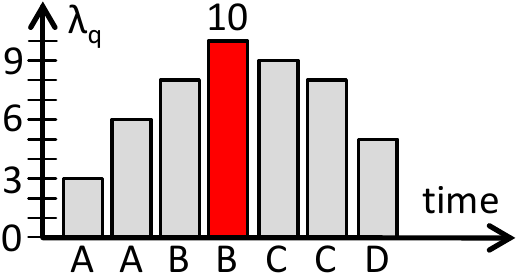}
  }
\ \  
 \subfloat[B-C: Latency peak.]{
    \label{fig:latency_queue_bc}
    \includegraphics[width=0.19\textwidth]{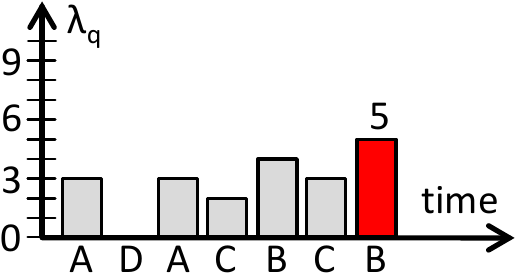}
  }
\ \  
 \subfloat[M-C: Latency peak.]{
    \label{fig:latency_queue_mc}
    \includegraphics[width=0.19\textwidth]{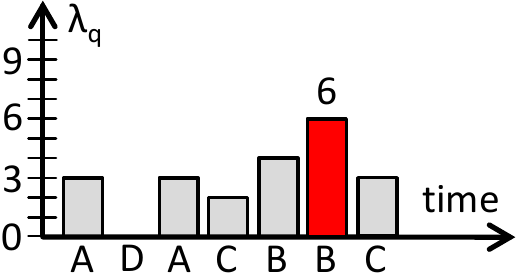}
  }
\vspace{-0.3cm}
\caption{Different sequences of negative and positive gains.}
\label{fig:gains}
\vspace{-0.35cm}
\end{figure*}

In data-parallel DCEP operators, in order to maintain the latency bound for each event, the batch scheduling controller decides at the \emph{start} of a window to which instance this window is scheduled. Then, it directs all events that arrive in the scope of that window to the corresponding instance. It is infeasible for the controller to wait until all events of the window are present and then schedule the window; it would take too much time in view of per-event latency bounds. 
After assigning a window to an operator instance at the occurrence of its start event, many other events of that window arrive until the window is finally closed. Thus, over the \emph{whole time span} of the window, feedback parameters in the operator instance are influenced by the scheduling decision, i.e., a long time after the scheduling decision has been made. That poses a completely different problem from other batch scheduling problems that are tackled with reactive batch scheduling, e.g., the problem of scheduling batches of events in streamed batch processing \cite{Das:2014:ASP:2670979.2670995}, where a controller \emph{first} builds a batch of available events and \emph{then} assigns it to an operator instance. 

Therefore, in data-parallel DCEP operators, there can be a high delay between assigning a window to an operator instance and the occurrence of the peak value of the feedback parameters in that operator instance. We denote this delay as the \emph{feedback delay}. In Figure \ref{table:1}, we have measured the feedback delay of operational latency and of queue length in the different runs of the traffic monitoring operator under different conditions; a feedback delay of 699 to 1,179 \emph{seconds} occurred for both parameters. In that time, many subsequent batch scheduling decisions have to be made by the controller. At the same time, key factors like inter-arrival time, window scope, event types, etc. continuously change. Moreover, the feedback delay is not constant, so that the controller cannot rely on it; it is not clear whether the parameter measured in an operator instance is already the peak value or how much further it will grow. 

To mitigate high feedback delays, we devise a \emph{latency-reactive} controller that reacts on the \emph{current} operational latency in operator instances. Windows are batched to the same operator instance until at the instance, the current operational latency reaches a threshold $\mathit{TH}$; subsequent windows are scheduled to the next operator instance. This, however, poses the question how to set $\mathit{TH}$. A simple experiment shows that a static $\mathit{TH}$ is not good enough to keep the latency bound $\mathit{LB}$. We run evaluations using the traffic monitoring operator at an average inter-arrival time of cars of 200 ms, aiming to keep $\mathit{LB} = 1s$. With $\mathit{TH} = 100 \mathit{ms}$, reactive batch scheduling more or less was able to keep $\mathit{LB}$ when $\mathit{ws}$ was not higher than 500 s (cf. Figure \ref{fig:eval/cumulated_latency_traffic_scenario}). However, at a $\mathit{ws}$ of 600 s and 700 s, $\mathit{TH} = 100 \mathit{ms}$ led to systematically wrong batch scheduling decisions; $\mathit{LB}$ was violated by a factor of almost 100. 
Obviously, $\mathit{TH}$ has to be adapted to the changing key factor values. In doing so, the feedback to change $\mathit{TH}$ is available only after $\mathit{LB}$ already has been violated, i.e., after a long feedback delay. 
The same problems apply when using the queue length peaks as a feedback parameter: The feedback delay is high. Again, using the current queue length as feedback parameter requires a suitable threshold, which in turn has to be adapted to changing key factor values. 

In the face recognition operator, window scopes are much smaller. While the feedback delay of operational latency peaks is still high (46 to 410 \emph{seconds}), the feedback delay of the queue length peaks is smaller (8 to 20 seconds; cf. Figure \ref{table:2}). However, this does not automatically make the queue length peaks a good parameter for reactive controllers. First of all, 20 seconds is still a long time; in the real-world workloads analyzed in Section \ref{sec:evaluation}, sudden bursts demand for an even faster reaction. Second, the relation between queue length peak and operational latency peak is not trivial; the operational latency peak does not necessarily occur when the most events are in the queue, but rather when the most expensive events are in the queue. This demands for a more thorough analysis. We conclude that neither operational latency nor queue length are a reliable feedback parameter for a purely reactive batch scheduling controller.

Instead of pure feedback mechanisms, our approach uses a simple, yet powerful latency model. It takes into account feedback from operator instances, but also includes a prediction and analysis step.

\section{Model-based Controller}
\label{sec:analytical}
The batch scheduling controller must predict whether the operational latency peak in an operator instance will be higher than $\mathit{LB}$ when batching a new window $\textsf{w}_{\mathit{new}}$. To this, we introduce a latency model. 
We aim to find the right balance between the complexity, the reasonable consideration of feedback from operator instances and of domain expert knowledge, and the accuracy and precision of the model.

\subsection{Basic Approach}
\label{sec:basic}

Recall that the operational latency of an event $e$ is built up of its queuing and processing latency: $\lambda_o(e) = \lambda_q(e) + \lambda_p(e)$. If the processing latency $\lambda_p(e)$ of an event is higher than the inter-arrival time $\mathit{iat}$ to its successor event, this imposes additional queuing latency to the successor event. On the other hand, if $\lambda_p(e)$ is smaller than $\mathit{iat}$, the queuing latency of the successor event becomes smaller or even zero, i.e., $e$ does not induce queuing latency for the successor event. In the following, we refer to the difference between $\lambda_p$ and $\mathit{iat}$ as the \emph{gain} $\gamma$ of an event: $\gamma(e) = \lambda_p(e) - \mathit{iat}$. If $\lambda_p(e) > \mathit{iat}$, we speak of a \emph{negative gain}; else, we speak of a \emph{positive gain}\footnote{Negative gains are positive numbers and positive gains are negative numbers. The terminology refers to the impact of an event on the feasibility to schedule a window in a batch.}. In Figure \ref{fig:latency_gains}, we provide an example. Suppose that the $\mathit{iat}$ between events is 5 time units (TU), and the window contains 7 events: 2 events of type A impose $\lambda_p$ = 8 TU, 2 events of type B impose $\lambda_p$ = 7 TU, 2 events of type C impose $\lambda_p$ = 4 TU, and 1 event of type D imposes $\lambda_p$ = 2 TU. Then, the gains of the single events are between +3 and -3 TU (+3 for type A, +2 for B, -1 for C, -3 for D).

Now, for the overall window $\textsf{w}_{\mathit{new}}$, the aggregated gains of the set of events with $\lambda_p(e) > \mathit{iat}$ are termed as the \emph{total negative gain}: $\Gamma^- = \sum \gamma(e) : e \in \textsf{w}_{\mathit{new}} \land \lambda_p(e) > \mathit{iat}$. In the given example (Figure \ref{fig:latency_gains}), those are the events of type A and B; hence, $\Gamma^- = 3 + 3 + 2 + 2 = 10$ TU. The aggregated gains of the set of events with $\lambda_p(e) < \mathit{iat}$ are termed as the \emph{total positive gain}\footnote{If $\lambda_p(e) = \mathit{iat}$, neither negative nor positive gains occur.}: $\Gamma^+ = \sum \gamma(e) : e \in \textsf{w}_{\mathit{new}} \land \lambda_p(e) < \mathit{iat}$.  In the given example (Figure \ref{fig:latency_gains}), those are the events of type C and D; hence, $\Gamma^+ = (-1) + (-1) + (-3) = -5 $ TU.

After defining the total negative and positive gains, in the following, we analyze possible sequences of negative and positive gains and the impact on the queuing latency peak $\lambda_q^{\mathit{max}}$. In Figure \ref{fig:latency_queue_wc}, first all negative gains occur, followed by all positive gains. This is the worst case with respect to $\lambda_q^{\mathit{max}}$; in the example sequence, $\lambda_q^{\mathit{max}}$ = 10 TU. Note, that also any other sequence of events of types A and B would lead to the same $\lambda_q^{\mathit{max}}$. In the worst case, hence, $\lambda_q^{\mathit{max}} = \Gamma^-$. However, an interleaving between negative and positive gains is possible as well. Take a look at Figures \ref{fig:latency_queue_bc} and \ref{fig:latency_queue_mc}: In the examples, the events with negative and positive gains interleave to a different extent. This leads to different values of $\lambda_q^{\mathit{max}}$, because although the queuing latency is increased by events with negative gains, events with positive gains compensate for that; a successor event of an event with positive gain faces a lower queuing latency.

The actual sequence of events with negative and positive gains in $\textsf{w}_{\mathit{new}}$ is very difficult to predict. It would essentially correspond to predicting each single event in $\textsf{w}_{\mathit{new}}$ and its $\mathit{iat}$. To account for the discussed interleaving of events with negative and positive gains, therefore, we introduce a \emph{compensation factor} $\alpha$. $\alpha$ allows for modeling the extent of interleaving of negative and positive gains without the need to explicitly define the sequence of events in $\textsf{w}_{\mathit{new}}$ in the prediction: $\lambda_q^{\mathit{max}} = \Gamma^- + \alpha * \Gamma^+$. Taking a look at the best-case example in Figure \ref{fig:latency_queue_bc}, we see that the negative and positive gains are maximally interleaving, hence, $\alpha = 1$. Accordingly, $\lambda_q^{\mathit{max}} = 10 + 1 * (-5) = 5$. Figure \ref{fig:latency_queue_mc} exemplifies an event sequence in between the worst- and best-case: Parts of the positive gains are interleaving with the negative gains, hence, $\alpha = 0.8$. Accordingly, $\lambda_q^{\mathit{max}} = 10 + 0.8 * (-5) = 6$.

Please notice, that the first event of $\textsf{w}_{\mathit{new}}$ might already face a queuing latency $\lambda_q^{\mathit{init}}$ at its arrival. This can be due to previous windows that had been scheduled to the same operator instance. Hence, the final formula to calculate the queuing latency peak is:\footnote{For the sake of readability, we did not mention in the text that $\lambda_q^{\mathit{max}} = \lambda_q^{\mathit{init}}$, if $\Gamma^- + \alpha * \Gamma^+ < 0$.} \\\centerline{$\lambda_q^{\mathit{max}} = \lambda_q^{\mathit{init}} + \Gamma^- + \alpha * \Gamma^+,\alpha \in [0,1]$.}

From the queuing latency peak $\lambda_q^{\mathit{max}}$, the operational latency peak $\lambda_o^{\mathit{max}}$ is calculated using the maximal processing latency $\lambda_p^{\mathit{max}}$ of any event in $\textsf{w}_{\mathit{new}}$. This bases on the pessimistic assumption that the most expensive event occurs right at the queuing latency peak; as we do not know the event sequence, this assumption is justified by the goal to avoid underestimations of $\lambda_o^{\mathit{max}}$. Hence, \\\centerline{$\lambda_o^{\mathit{max}} = \lambda_q^{\mathit{max}} + \lambda_p^{\mathit{max}}$.}

Using this latency model, the operational latency peak can be predicted, and the scheduling decision---to batch or not to batch---can be made accordingly. In the following sub-section, we describe how the parameters of the model are predicted.

\begin{figure}
\begin{minipage}[t]{0.48\linewidth}
    \includegraphics[width=\linewidth]{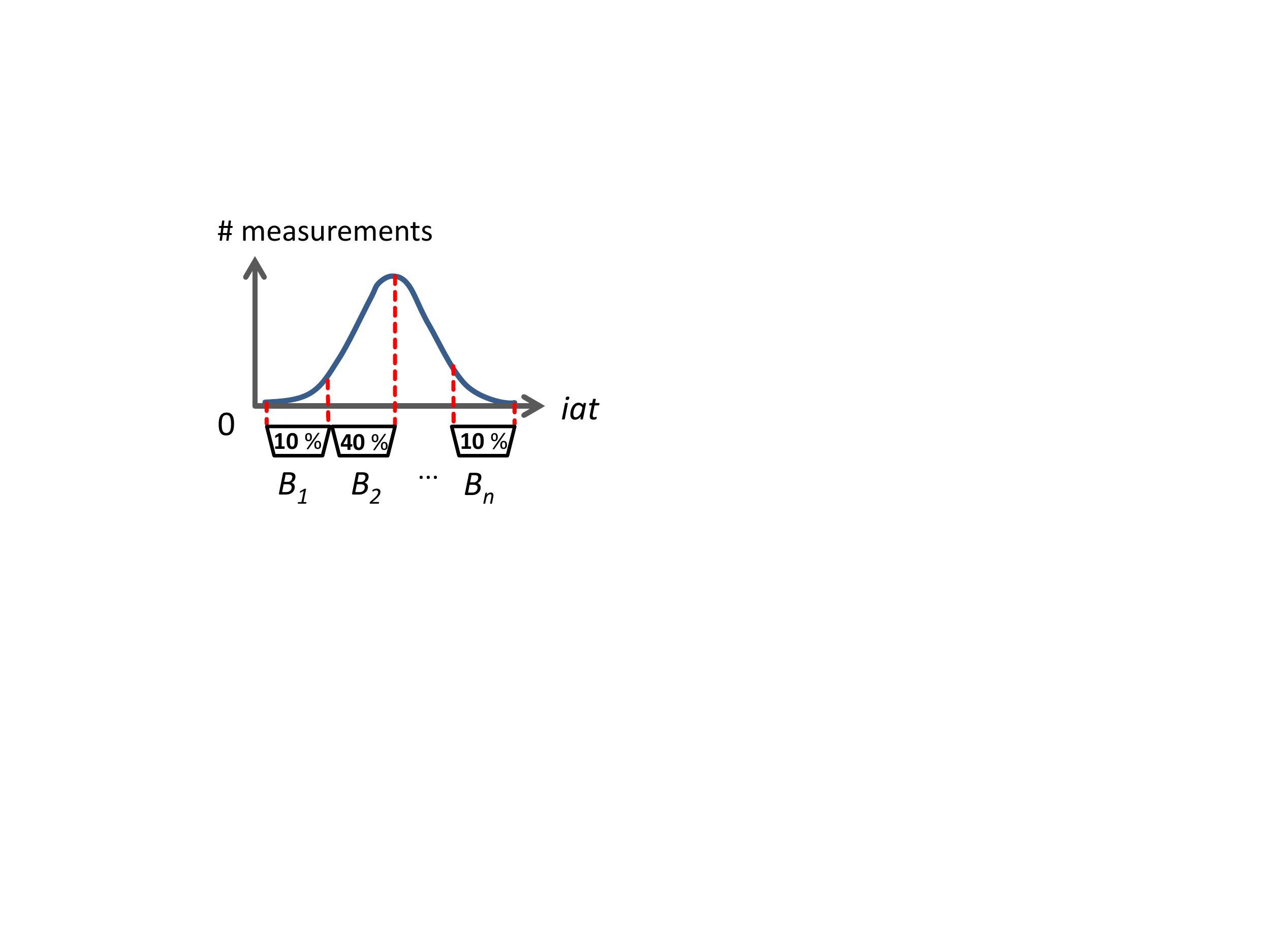}
\vspace{-0.6cm}
    \caption{\textit{iat} bins.}
    \label{fig:bins}
\end{minipage}%
    \hfill%
\begin{minipage}[t]{0.33\linewidth}
    \includegraphics[width=\linewidth]{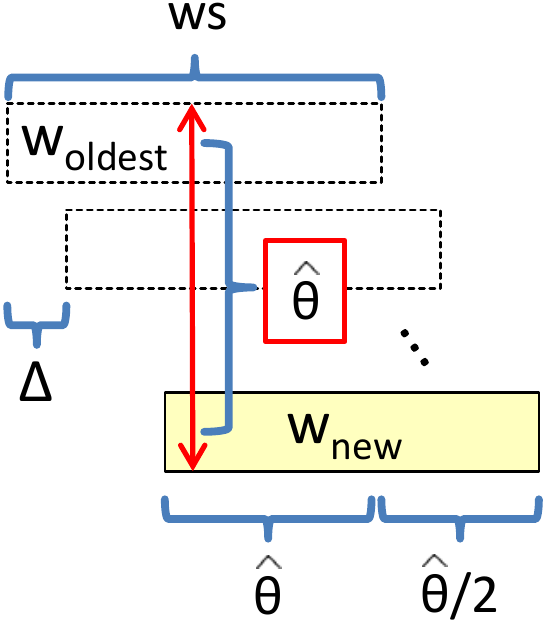}
\vspace{-0.6cm}
    \caption{Overlap.}
    \label{fig:overlap_prediction}
\end{minipage}%
\vspace{-0.5cm}
\end{figure}

\subsection{Prediction of Model Parameters}
\label{sec:prediction}

The proposed latency model builds on predicting the total sum of negative and positive gains of all events in $\textsf{w}_{\mathit{new}}$; i.e., it does not regard individual events, but it regards events in $\textsf{w}_{\mathit{new}}$ as \emph{sets} of events imposing negative or positive gains. Hence, it builds on the prediction of the \emph{set} of events in $\textsf{w}_{\mathit{new}}$, including their processing latency $\lambda_p$ and their inter-arrival time $\mathit{iat}$. Further, a prediction of the initial queuing latency $\lambda_q^{\mathit{init}}$ and the compensation factor $\alpha$ is needed. Based on those values, the model predicts the operational latency peak.
In this section, we discuss appropriate prediction methods and algorithms. 

\textbf{Inter-arrival time}. The splitter continuously monitors the past $\mathit{iat}$ values in a window of $\mathit{mtime}$ time units. Our $\mathit{iat}$ model tackles two challenges: heavy fluctuations of the $\mathit{iat}$ around an average value (variance) and rapid changes of the average $\mathit{iat}$ (changing trend). 

Tackling the first challenge, the splitter arranges the monitored inter-arrival times in a discrete model (cf. Figure \ref{fig:bins}). The range of measured $\mathit{iat}$ values is divided into a number of equally-sized \emph{bins}. The measured $\mathit{iat}$s are sorted into the corresponding bin; for each bin $B_i$, the mean value $\overline{\mathit{iat(B_i)}}'$ is computed: $\mathit{iat(B_i)} = \overline{\mathit{iat(B_i)}}'$. Each bin is assigned with a $\mathit{weight(B_i)}$, i.e. ratio of number of entries in the bin to total number of measurements in all bins. The number of bins manages the accuracy of the model; the minimum number of bins is 1.

Tackling the second challenge, we introduce a negative bias on the monitored mean value $\overline{\mathit{iat(B_i)}}'$ in each bin. This way, the model accounts for changes in the average $\mathit{iat}$ between the monitored value $\overline{\mathit{iat(B_i)}}'$ and the value that will occur in $\textsf{w}_{\mathit{new}}$. The negative bias is modeled based on a factor $\delta_{\mathit{iat}}$ of standard deviations $\sigma$ of the monitored $\mathit{iat}$s, e.g., 1 standard deviation or 2 standard deviations. Then, $\mathit{iat(B_i)} = \overline{\mathit{iat(B_i)}}' - \delta_{\mathit{iat}} * \sigma$. 

\textbf{Processing Latency.} In our model, $\lambda_p$ depends on the overlap $\Theta$ and the processing latency in a single window $\lambda_p^\textsf{w}$: $\lambda_p = \Theta * \lambda_p^\textsf{w}$. As discussed in Section \ref{sec:problemdescription}, $\lambda_p^\textsf{w}$ depends on the event type and the position in a window. Hence, first of all, our model differentiates between different event types. This design decision has two consequences: First, the prediction model of $\lambda_p^\textsf{w}$ takes into account the type, i.e., predict $\lambda_p^\textsf{w}(\mathit{type})$, the in-window processing latency of events of a specific type. Second, the set of events in $\textsf{w}_{\mathit{new}}$ is predicted with respect to the number of events of different types.

For modeling  $\lambda_p^\textsf{w}({\mathit{type}})$, we propose the same methods as for modeling $\mathit{iat}$,  using a combination of negative bias and bins. Same as in $\mathit{iat}$ bins, in each latency bin $B_l$, we predict $\mathit{\lambda_p^\textsf{w}(B_l)} = \overline{\mathit{\lambda_p^\textsf{w}(B_i)}}' + \delta_{\lambda_p} * \sigma$, i.e., the measured mean in-window processing latency in the latency bin plus a factor $\delta_{\lambda_p}$ of standard deviations.
The advantage of monitoring the current (distribution of) $\lambda_p^\textsf{w}({\mathit{type}})$ in the operator instances over building a position-dependent latency model is that we can \emph{implicitly} incorporate the \emph{position-dependency}: When the (distribution of) positions of events in windows change, e.g., due to changing workload or changing window scopes, this is reflected in the monitored current (distribution of) $\lambda_p^\textsf{w}({\mathit{type}})$ values. We do not need to explicitly model the positions of individual events. 

The overlap $\Theta$ for all events of $\textsf{w}_{\mathit{new}}$ is modeled as the average overlap of events of $\textsf{w}_{\mathit{new}}$ in the current batch, denoted by $\overline{\Theta}$. 
Predicting $\overline{\Theta}$ is performed according to the following model (cf. Figure \ref{fig:overlap_prediction}). When $\textsf{w}_{\mathit{new}}$ is scheduled in a batch of already opened windows, a number of events in $\textsf{w}_{\mathit{new}}$ has the current overlap $\hat{\Theta}$, until the oldest open window $\textsf{w}_{\mathit{oldest}}$ in the batch closes. From closing $\textsf{w}_{\mathit{oldest}}$ until closing $\textsf{w}_{\mathit{new}}$, the overlap decreases step-wise in regular intervals each time a window between $\textsf{w}_{\mathit{oldest}}$ and $\textsf{w}_{\mathit{new}}$ is closed. In that phase, the average overlap is $\hat{\Theta} / 2$. 
In order to compute $\overline{\Theta}$, we weigh the ratio of events with overlap $\hat{\Theta}$ to the events with overlap $\hat{\Theta} / 2$. In doing so, we assume in our model that all windows in the batch have the same window scope $\mathit{ws}$, and between the start of two windows there is the same shift $\Delta$; $\mathit{ws}$ and $\Delta$ are measured in the splitter at regular intervals to keep them up to date at each scheduling decision.

At the start of $\textsf{w}_{\mathit{new}}$, $\textsf{w}_{\mathit{oldest}}$ is already open since $(\hat{\Theta} - 1) * \Delta$ time units, as $\hat{\Theta} - 1$ is the number of windows between $\textsf{w}_{\mathit{oldest}}$ and $\textsf{w}_{\mathit{new}}$ that were opened in intervals of $\Delta$ time units. Therefore, $\textsf{w}_{\mathit{oldest}}$ stays open for $\mathit{ws} - (\hat{\Theta} - 1) * \Delta$ more time units. When $\textsf{w}_{\mathit{oldest}}$ closes, the phase of closing windows starts, spanning $(\hat{\Theta} - 1) * \Delta$ time units. Hence, the weighed average overlap is computed as follows:
\vspace{-0.1cm}
\begin{center}
$\overline{\Theta} = \frac{(\mathit{ws} - (\hat{\Theta} - 1) * \Delta) * \hat{\Theta} +  (\hat{\Theta} - 1) * \Delta * \hat{\Theta}/2}{\mathit{ws}}$. 
\end{center}
\vspace{-0.1cm}

\textbf{Number of Events.}
For predicting the set of events in $\textsf{w}_{\mathit{new}}$, there are three significant factors in the model: (1) The window scope $\mathit{ws}$, (2) the $\mathit{iat}$, and (3) the ratio of different event types, denoted as $\mathit{ratio(type)}$, that models which percentage of events in $\textsf{w}_{\mathit{new}}$ is of a specific $\mathit{type}$. These factors are gained from monitoring them in the incoming event stream in the splitter in the past $\mathit{mtime}$ time units. To predict the total number of events in $\textsf{w}_{\mathit{new}}$, we again use a negative bias of $\delta_{\mathit{iat}}$ standard deviations $\sigma(\mathit{iat})$, so that $\mathit{iat} = \overline{\mathit{iat}}' - \delta_{\mathit{iat}} * \sigma(\mathit{iat})$. Then, the total number of events $n$ is predicted as $n = \frac{\mathit{ws}}{\mathit{iat}}$, and the number of events of a specific $\mathit{type}$, denoted by $\#(\mathit{type})$, is predicted as $\mathit{ratio(type)} * n$.

\textbf{Initial Queuing Latency.}
The initial queuing latency is predicted for each operator instance separately, depending on the content of the incoming event queue. To this end, operator instances report the number of events of each type and their average overlap $\overline{\Theta}$ in the assigned windows in regular intervals to the splitter. The splitter calculates $\lambda_q^{\mathit{init}}$ of an operator instance as the sum of the processing latencies of all reported events in its queue: $\lambda_q^{\mathit{init}} = \sum_{\mathit{types}}\# \mathit{events} * \overline{\Theta} * \lambda_p^{\textsf{w}}({\mathit{type}})$.

\begin{figure}
\setbox0\vbox{\small
{\fontsize{6.5}{7.5}\selectfont
\begin{algorithmic}[1]
\algsetblockdefx[function]{func}{endfunc}{}{0.2cm}[3]{#1 \textbf{#2} (#3) \textbf{begin}}{\textbf{end function}}

\func{$\langle \mathit{long, long} \rangle$}{predictGains}{ } \Comment{returns $\Gamma^-$ and $\Gamma^+$}
	
	\State \texttt{predict $\#\mathit{events}$ for each latency bin $B_l$: $\#(B_l)$}
	\State \texttt{sort latency bins by mean latency (highest first)}
	\State \texttt{predict $\#\mathit{events}$ for each iat bin $B_i$:  $\#(B_i)$}
	\State \texttt{sort iat bins by mean iat (lowest first)}
	\While{\texttt{true}}
		\State $\mathit{\#combination} \gets min\{ \#(B_l), \#(B_i) \} $
		
		\State $\mathit{gain} \gets \mathit{\#combination} * (\overline{\Theta} * \lambda_p^\textsf{w}(B_l) - \mathit{iat(B_i)})$
		\If{$\mathit{gain} > 0$}
			\State $\Gamma^- \gets \Gamma^- + \mathit{gain}$ 
		\Else
			\State $\Gamma^+ \gets \Gamma^+ + \mathit{gain}$ 
		\EndIf
		
		\State $\#(B_l) \gets \#(B_l) - \mathit{\#combination}$
		\State $\#(B_i) \gets \#(B_i) - \mathit{\#combination}$

		\If{$\#(B_i) = 0$}
			\State $i \gets i + 1$ \Comment{next iat bin} 
		\EndIf
		\If{$\#(B_l) = 0$}
			\State $l \gets l + 1$ \Comment{next latency bin} 
		\EndIf
		
		\If{\texttt{no more bins}}
			\State \textbf{return} $\langle \Gamma^-,\Gamma^+ \rangle$
		\EndIf
	\EndWhile
\endfunc

\end{algorithmic}
}
}
\centerline{\fbox{\box0}}
\vspace{-0.25cm}
\caption{Predict negative and positive gains.}
\vspace{-0.3cm}
\label{fig:gainscalc}
\end{figure}

\textbf{Compensation Factor.}
For modeling the compensation factor $\alpha$, there are two possibilities.

First, we propose a heuristic, denoted as T-COUNT, for adapting $\alpha$ based on the current extent of interleaving between events with different processing latency in the incoming stream. To this end, events are divided into two groups, based on their in-window processing latency $\lambda_p^\textsf{w}$: the group of events with higher $\lambda_p^\textsf{w}$ is denoted by $T^-$ and the group of events with lower $\lambda_p^\textsf{w}$ is denoted by $T^+$. The distinction between the groups is made based on the average $\lambda_p^\textsf{w}(\mathit{type)}$ of the event types; there is one half of event types that has higher $\lambda_p^\textsf{w}(\mathit{type})$ than the other half of event types. Events of any of the types that pose higher processing latencies are grouped into $T^-$, other events are grouped into $T^+$. The splitter continuously counts in a monitoring window of temporal size $\mathit{mtime}$, how many events in $T^-$, denoted by $c^-$, and how many events in $T^+$, denoted by $c^+$, occur. Further, the splitter counts how often events in $T^-$ and $T^+$ follow each other, i.e., the number of transitions, denoted by $c^t$. The maximal number of transitions is $2 * \mathit{min}\{c^+, c^- \}$. Trivially, the minimum number of transitions is 1. Then, $\alpha$ is predicted as the proportion of $c^t$ to the maximal number of transitions: $\alpha = \frac{c^t - 1}{2 * \mathit{min}\{c^+, c^- \}}$. 

Second, a domain expert can also set a fixed or dynamic value of $\alpha$ based on off-line training if the characteristics of the expected workloads are known beforehand.

\begin{figure}
\setbox0\vbox{\small
{\fontsize{6.5}{7.5}\selectfont
\begin{algorithmic}[1]
\algsetblockdefx[function]{func}{endfunc}{}{0.2cm}[3]{#1 \textbf{#2} (#3) \textbf{begin}}{\textbf{end function}}

\State OperatorInstance $\omega_x$ \Comment{current operator instance}

\func{void}{schedule}{ }

	\State $\lambda_o^\mathit{max}$ $\gets$ $\mathit{LatencyModel.newPrediction()}$ 
		
	\If{$\lambda_o^\mathit{max} \leq \mathit{LB}$}
		\State assign $\sigma$ to $\omega_x$ 
	\Else
		\State $x \gets (x + 1)$ MOD $\#op\_instances$ \Comment{Round-Robin}
		\State assign $\sigma$ to $\omega_x$
	\EndIf

\endfunc

\end{algorithmic}
}
}
\centerline{\fbox{\box0}}
\vspace{-0.2cm}
\caption{Scheduling algorithm.}
\label{fig:salg}
\vspace{-0.5cm}
\end{figure}

\subsection{Scheduling Algorithm}

Having a prediction of the set of events in $\textsf{w}_{\mathit{new}}$, processing latencies and inter-arrival times, the batch scheduling controller predicts the total negative and positive gains and the operational latency peak in order to schedule $\textsf{w}_{\mathit{new}}$. In this section, we introduce the algorithms.

\textbf{Total Negative and Positive Gains Prediction.}
To predict $\Gamma^-$ and $\Gamma^+$, the predicted processing latencies and inter-arrival times have to be combined. Each processing latency bin represents a number of events in $\textsf{w}_{\mathit{new}}$ having a specific $\lambda_p$; each $\mathit{iat}$ bin represents a number of events having a specific $\mathit{iat}$. In order to calculate the total negative and positive gain of all events, the number of events having a specific combination of $\lambda_p$ and $\mathit{iat}$ is predicted. To this end, events from the bin with highest $\lambda_p$ are combined with the lowest $\mathit{iat}$, etc., and events with lowest $\lambda_p$ are combined with the highest $\mathit{iat}$. The concrete algorithm is presented in the following (cf. algorithm in Figure \ref{fig:gainscalc}).
First, for each type, the total number of events, $\#(\mathit{type})$, is divided into latency bins according to the weights of the bins: The number of events $\#(B_l)$ in a latency bin $B_l$ is: $\#(B_l) = \#(\mathit{type}) * \mathit{weight}(B_l)$. Then, all latency bins of all event types are globally sorted by their mean processing latency (highest first). The $\mathit{iat}$ bins are sorted by the mean $\mathit{iat}$ (lowest $\mathit{iat}$ first); the number of events $\#(B_i)$ in an $\mathit{iat}$ bin $B_i$ is computed based on the total number of events, $n$, and the weight of the bin, $\#(B_i) = n * \mathit{weight}(B_i)$. Then, the numbers of events in the processing latency bins and $\mathit{iat}$ bins are combined such that the highest processing latencies are combined with the lowest $\mathit{iat}$s. The algorithm iterates through the bins (lines 6 -- 25): For the combination of current latency and $\mathit{iat}$ bin, the gain of the events in this combination is calculated based on the processing latency and the $\mathit{iat}$ of the bins. If the predicted gain is greater than 0, it is added to the total negative gains, else it is added to the total positive gains. Then, the next combination of bins is processed. When the iteration went through all bins, the resulting total negative and positive gains are returned.

\textbf{Operational Latency Peak.}
The operational latency peak $\lambda_o^{\mathit{max}}$ is predicted with the formulas introduced in Section \ref{sec:basic}, taking into account the predicted parameters as described in Section \ref{sec:prediction}: $\lambda_o^{\mathit{max}} = \lambda_q^{\mathit{max}} + \lambda_p^{\mathit{max}}$, with $\lambda_q^{\mathit{max}} = \lambda_q^{\mathit{init}} + \Gamma^- + \alpha * \Gamma^+$. In doing so, $\lambda_p^{\mathit{max}}$ is predicted as the in-window processing latency $\lambda_p^\textsf{w}$ of the most expensive event type in the most expensive latency bin, denoted as $\mathit{max}(\lambda_p^\textsf{w})$, at the average overlap: $\lambda_p^{\mathit{max}} = \overline{\Theta} * \mathit{max}(\lambda_p^\textsf{w})$.

\textbf{Scheduling.}
When scheduling a new window, the controller checks whether batching it to the same operator instance where the last window was assigned to would lead to a violation of $\mathit{LB}$. The scheduling algorithm is listed in Figure \ref{fig:salg}. The latency model is queried for a prediction of the operational latency peak $\lambda_o^\mathit{max}$ (line 3). The predicted $\lambda_o^\mathit{max}$ is compared to $\mathit{LB}$ and a batch scheduling decision is made accordingly: If $\lambda_o^\mathit{max} \leq \mathit{LB}$, the window is assigned to the same instance as the last window (lines 4--5); else, it is scheduled to the next operator instance according to the Round-Robin algorithm (lines 6--8).

\section{Evaluation}
\label{sec:evaluation}

In our evaluations, we analyze the proposed methods in two steps. In a first step, we perform a distinct evaluation of the proposed latency model. We show the accuracy and precision of the latency model in predicting the negative gains, positive gains and latency peaks in different situations under synthetic workloads.  In the second step, we measure the performance of the overall event processing system under different realistic conditions---such as inter-arrival times and latency bounds---comparing the model-based batch scheduling controller to Round-Robin and to a reactive batch scheduling algorithm. The cost of prediction is also evaluated.

\begin{figure}
\footnotesize
\begin{tabular}{| c | p{6.5cm} |} \hline
  Symbol & Parameter Description \\ \hline\hline
  $\mathit{iat}$ & average inter-arrival time of events \\ \hline
  $b$ & batch size, i.e., number of subsequent  windows scheduled to same op. instance  \\ \hline
  $\mathit{ws}$ & window scope, i.e., temporal scope of a window  \\ \hline
  $\Gamma^-, \Gamma^+$ & total negative and positive gains \\ \hline
  $\alpha$ & compensation factor \\ \hline
  $\lambda_o$, $\lambda_q$, $\lambda_p$ & operational latency, queuing latency and processing latency of an event in an operator instance; $\lambda_o = \lambda_q + \lambda_p$  \\ \hline
  $\lambda_q^{\mathit{max}}$ & queuing latency peak:  $\lambda_q^{\mathit{max}} = \lambda_q^{\mathit{init}} + \Gamma^- + \alpha * \Gamma^+$\\ \hline
 $\lambda_q^{\mathit{init}}$ & initial queuing latency before processing the first event of a window   \\ \hline
  $\mathit{LB}$ & latency bound, i.e., the peak operational latency that shall not be exceeded\\ \hline
  $RR$ & Round-Robin scheduling, circularly assigns one window to each operator instance \\ \hline
  $\delta_{\mathit{iat}}, \delta_{\lambda_p}$ & negative bias of measured $\mathit{iat}$ or $\lambda_p$ in the monitoring window, in std. deviations: e.g., $\mathit{iat} - \delta_{\mathit{iat}} * \sigma$ \\ \hline
  $\mathit{mtime}$ & size of the workload monitoring window  \\ \hline
  $\mathit{TH}$ & scheduling threshold of reactive baseline controller, cf. Section \ref{sec:reactive} \\ \hline
\end{tabular}
\vspace{-0.1cm}
\caption{Symbols used.}
\label{fig:symbols}
\vspace{-0.4cm}
\end{figure}

\textbf{Experimental Setup and Notation.}
To evaluate the batch scheduling controller, we have integrated it into an existing data parallelization framework \cite{7024105}.
All experiments were performed on a computing cluster consisting of 16 homogeneous hosts with each 8 CPU cores (Intel(r) Xeon(R) CPU E5620 @ 2.40 GHz) and 24 GB memory, connected by 10-GB Ethernet links. The components of the parallelization framework were distributed among the available hosts.
Symbols used in the evaluations are listed in Figure \ref{fig:symbols}.

\subsection{Latency Model}
\label{sec:ev_latencymodel}

In the following, we evaluate the accuracy and precision of the proposed latency model. We present the evaluation in two parts: First, we evaluate the predictions of the total negative and positive gains. Based on that, we then analyze the prediction of the queuing latency peak, which depends on the prediction of negative and positive gains as well as on the compensation factor $\alpha$.

\begin{figure}
\centering
\begin{minipage}[t]{0.62\linewidth}
    \includegraphics[width=\linewidth]{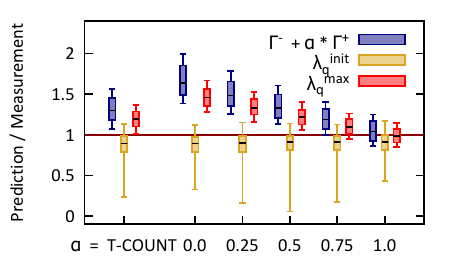} 
    \label{fig:video_lambda_q_batching_size_4_ps_10s}    
 \end{minipage}
\vspace{-0.6cm}
\caption{Predictions of queuing latency peak. Face recognition operator, $b = 4$, $\mathit{ws} = 10 s$. }
\label{fig:alpha}
\vspace{-0.5cm}
\end{figure}

\textbf{Interpretation of the figures in this section.}
We measured both the predicted values as well as the values that actually occurred in the operator instances. In all experiment results, i.e., Figure \ref{fig:eval_gain} and Figure \ref{fig:alpha}, on the y-axis, we depict the predicted values normalized to the measured values. For example, a value of 1.0 means that the prediction exactly met the actually occurred value, a value smaller than 1.0 means that the prediction was too low (i.e., underestimation), and a value higher than 1.0 means that the prediction was too high (i.e., overestimation). All figures depict the 10th, 25th, 50th, 75th, and 90th quantiles in a ``candlesticks'' representation.

\subsubsection{Negative and Positive Gains}
\label{sec:Negative and Positive Gains}
In analyzing the prediction of $\Gamma^-$ and $\Gamma^+$, we run evaluations on synthetic workloads. Using synthetic workloads allows us to perform measurements in controlled situations where all of the parameters are well-known and completely under our control. This is not the case in real-world workloads, as we use them in the analysis of the overall event processing system in Section \ref{sec:performance}.
For the face recognition operator, we created a synthetic stream of face events (i.e. images containing a person's face). Each 2 seconds, a burst of 4 face events with an inter-arrival time of 10 ms was created, which resembles 4 persons in front of a camera that captures a picture each 2 seconds. The query events were generated with a fixed rate of 1 query per second, so that each second, one new window was started.  For the traffic monitoring operator, we created a workload trace with an average inter-arrival time of events of 100 ms following an exponential distribution, which resembles 5 cars per second passing each road checkpoint.

\begin{figure}
\begin{minipage}[t]{0.5\linewidth}    
	\begin{overpic}[width=\linewidth]{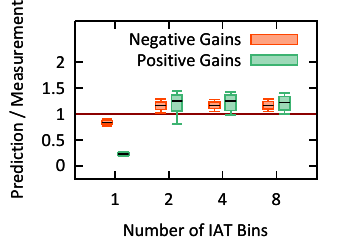}
        		\put(0,3){(a)}
      	\end{overpic}
 \end{minipage}
\begin{minipage}[t]{0.49\linewidth}
	\begin{overpic}[width=\linewidth]{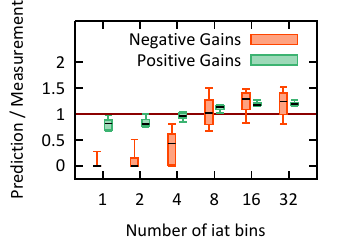}
        		\put(0,3){(b)}
      	\end{overpic}
\end{minipage}
\vspace{-0.6cm}
\caption{Prediction of negative and positive gains. (a) Face recognition operator, $b = 4$, $\mathit{ws} = 10 s$, (b) Traffic monitoring operator, $b = 1000$, $\mathit{ws} = 500 s$. }
\label{fig:eval_gain}
\vspace{-0.5cm}
\end{figure}

Figure \ref{fig:eval_gain}a shows evaluations of the face recognition operator using a different number of $\mathit{iat}$ bins. If only 1 bin is used, the predictions of $\Gamma^-$ and $\Gamma^+$ are poor. With a growing number of $\mathit{iat}$ bins, the latency model becomes more accurate: With 2, 4 or 8 bins, the predictions of both  $\Gamma^-$ and $\Gamma^+$ are very accurate and precise. 
In the traffic monitoring operator, employing more $\mathit{iat}$ bins, as shown in the results in Figure \ref{fig:eval_gain}b, quickly improves the prediction quality as well. 

We have also evaluated the effects of a negative bias on $\mathit{iat}$, as well as bins and negative bias on processing latency; a detailed discussion with all the results can be found in \cite{TR-2016-04}. 
Summarizing those results, using a negative bias of $\delta_{\mathit{iat}}$ standard deviations in the $\mathit{iat}$ bins makes the model more pessimistic. Further, the negative and positive gains in the tested scenarios are dominated by $\mathit{iat}$ such that employing bins and negative bias only on processing latency is insufficient to tune the model.

\subsubsection{Queuing Latency Peak}
Recall that the queuing latency peak is predicted based on the total negative and positive gains and the compensation factor $\alpha$: $\lambda_q^{\mathit{max}} = \lambda_q^{\mathit{init}} + \Gamma^- + \alpha * \Gamma^+$.
We show on the example of the face recognition operator that our proposed T-COUNT heuristic provides a suitable, slightly pessimistic estimation of $\alpha$ such that no under-estimation of queuing latency peak occurs. Additionally, we evaluate the prediction of the initial queuing latency $\lambda_q^\mathit{init}$. Following our observations from Section \ref{sec:Negative and Positive Gains}, we employ the latency model with 2 $\mathit{iat}$ bins, so that the predictions of $\Gamma^-$ and $\Gamma^+$  are accurate.

We see in  Figure \ref{fig:alpha} that the T-COUNT heuristics leads to a good overall estimation of $\lambda_q^{\mathit{max}}$. In predicting $\lambda_q^\mathit{init}$, fluctuations are caused by events in the network that have not yet arrived in the queue of an operator instance and are not considered in the feedback to the splitter. However, the impact of this behavior on the prediction of $\lambda_q^{\mathit{max}}$ is small, as $\lambda_q^{\mathit{max}}$ is dominated by the negative and positive gains. 

Besides the T-COUNT heuristic, we also systematically evaluated the impact of fixed values of $\alpha$ on the prediction of $\lambda_q^{\mathit{max}}$. As can be seen in Figure \ref{fig:alpha}, using different fixed values leads to different degrees of over- or underestimations of $\lambda_q^{\mathit{max}}$. Off-line profiling can be used in order to develop optimally pessimistic or optimistic models to set $\alpha$, when the characteristics of the workload are well-known before system deployment.

\subsection{Overall Event Processing System}
\label{sec:performance}

We compare our model-based batch scheduling controller to two baseline scheduling algorithms: Round-Robin scheduling and latency-reactive scheduling. Round-Robin aims for good load balancing but disregards communication overhead; it is the standard scheduling algorithm used in window-based data parallelization systems such as \cite{7024105}. Latency-reactive scheduling, as described in \ref{sec:reactive}, batches windows to an operator instance until its operational latency exceeds a threshold $\mathit{TH}$. It is used as a latency-aware baseline algorithm to compete against our model-based controller.

\begin{figure}
\begin{minipage}[t]{0.54\linewidth}
	\begin{overpic}[width=\linewidth]{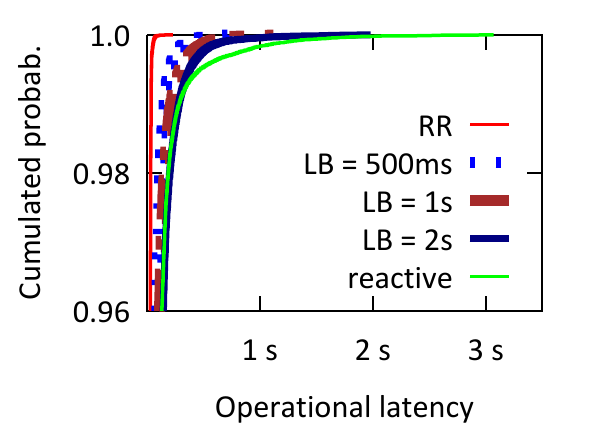}
        		\put(0,0){(a)}
      	\end{overpic}
\end{minipage}%
    \hfill%
\begin{minipage}[t]{0.45\linewidth}
	\begin{overpic}[width=\linewidth]{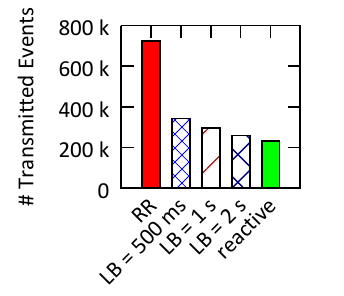}
        		\put(0,0){(b)}
      	\end{overpic}
\end{minipage}%
    \hfill%
\vspace{-0.2cm}
\caption{Traffic monitoring operator. (a) Operational latency. (b) Communication cost. }
\label{fig:evalTM}
\vspace{-0.5cm}
\end{figure}

\textbf{Traffic Monitoring Scenario.} 
In our dynamic traffic monitoring scenario, we modeled the inter-arrival time of vehicles as an exponential distribution with an average value following a sinusoidal curve between 2000 ms and 200 ms. 
Following the evaluation of the latency model in Section \ref{sec:ev_latencymodel}, we set-up the controller to use 8 $\mathit{iat}$ bins and a tumbling monitoring window with  $\mathit{mtime} = 60 s$. To account for the position-dependency of the operator and the rapdily changing workload, we add a pessimistic bias of $\delta_{\lambda_p} = 2$ standard deviations on the monitored processing latency and $\delta_{\mathit{iat}} = 0.75$ standard deviations on the monitored $\mathit{iat}$.
In all experiments, the parallelization degree, i.e., number of operator instances, was fixed at 8. Each experiment was running for 5 hours. 

At a window scope of 500 seconds, Round-Robin scheduling resulted in a maximal operational latency of 200 ms (cf. Figure \ref{fig:evalTM}a) and 724,464 events have been transmitted between the splitter and the operator instances (cf. Figure \ref{fig:evalTM}b). We ran the same experiment using our batch scheduling controller allowing for 2.5, 5 and 10 times higher operational latency peaks than yielded in Round-Robin: 500 ms, 1 s and 2 s. As shown in Figure \ref{fig:evalTM}a, $\mathit{LB}$ was kept. \textbf{\emph{The communication overhead was reduced by 53 \%, 59 \% and 64 \%, respectively}} (cf. Figure \ref{fig:evalTM}b). We compared this performance to the latency-reactive scheduler described in Section \ref{sec:reactive}; the reactive scheduler batches windows to an operator instance until it reports a current operational latency of more than $\mathit{TH} = 100ms$. The operational latency and communication overhead was very similar to model-based scheduling at $\mathit{LB} = 2 s$; however, the tail of the latency distribution is much longer, leading to 50 \% higher operational latency peaks. This indicates that the reactive scheduler erratically batches too many windows, leading to a less predictable behavior of the operator instances than when the model-based controller is used.

\textbf{Face Recognition Scenario.}
With the dynamic face recognition scenario, we evaluate the system behavior at a highly bursty \emph{real-world workload.}
A \emph{real video stream} from a camera installed on campus---capturing 1 frame each 2 seconds---is processed by a face detection operator and the detected faces are streamed to the face recognition operator. Simulating users of a face recognition application, the arrival of new queries is modeled as an exponential distribution with an average inter-arrival time of 2 seconds. The face recognition operator detects whether the queried person is in the face event stream, using a window scope of $\mathit{ws} = 10 s$. Each experiment ran for 150 minutes. According to the insights we gained from the evaluation of the latency model in Section \ref{sec:ev_latencymodel}, we set-up the controller to use 2 $\mathit{iat}$ bins. Further, we set $\mathit{mtime} = 10s$ (tumbling window) and $\delta_{\mathit{iat}} = 1.0$ standard deviations to account for the changing $\mathit{iat}$. 

\begin{figure}
\begin{minipage}[t]{0.54\linewidth}
	\begin{overpic}[width=\linewidth]{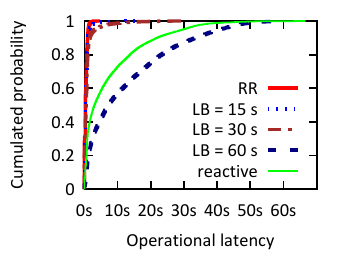}
        		\put(0,5){(a)}
      	\end{overpic}
    \label{fig:cumulated_latency_video_scenario_predictive}
\end{minipage}%
    \hfill%
\begin{minipage}[t]{0.45\linewidth}
	\begin{overpic}[width=\linewidth]{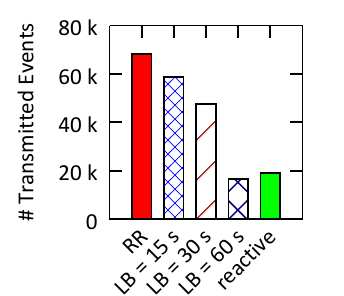}
        		\put(0,5){(b)}
      	\end{overpic}
    \label{fig:events_transmitted_video}
\end{minipage}%
    \hfill%
\vspace{-0.7cm}
\caption{Face recognition operator. (a) Operational latency. (b) Communication cost.}
\label{fig:eval_face}
\vspace{-0.5cm}
\end{figure}

For Round-Robin scheduling, we measured an operational latency peak of 6 seconds (cf. Figure \ref{fig:eval_face}a) and 68,412 events have been transmitted between the splitter and the operator instances (cf. Figure \ref{fig:eval_face}b). We ran the same experiment using our batch scheduling controller allowing for 2.5, 5 and 10 times higher operational latency peaks than yielded in Round-Robin: 15 s, 30 s and 60 s.
The latency bounds are kept in all tested settings (cf. Figure \ref{fig:eval_face}a).  \textbf{\emph{The communication overhead was reduced by 14 \%, 31 \% and 76 \%, respectively}} (cf. Figure \ref{fig:eval_face}b). We compared this performance to the latency-reactive scheduler described in Section \ref{sec:reactive} with $\mathit{TH} = 6s$. The operational latency peaks were 15 \% higher than with the model-based controller at $\mathit{LB} = 60 s$, while the communication overhead was 14 \% higher as well. With a higher threshold $\mathit{TH}$, the reactive scheduler would induce even higher latency peaks, while with a lower  $\mathit{TH}$, it would induce an even higher communication overhead; hence, the model-based controller is more effective, no matter how the reactive scheduler's threshold is set up.

In summary, model-based batch scheduling is effective in trading communication overhead against operational latency. In comparison, reactive scheduling is less predictable and effective than model-based scheduling; it might still be useful in cases where a simple best-effort batching approach is sufficient, but should not be used when latency bounds must be enforced.

\begin{figure}[h]
\begin{minipage}[t]{0.41\linewidth}
	\begin{overpic}[width=\linewidth]{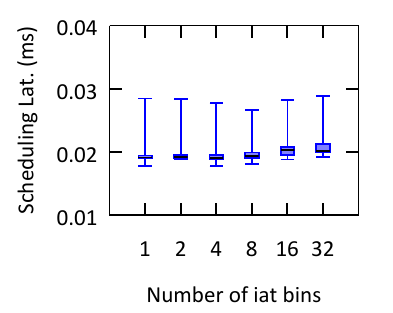}
        		\put(0,5){(a)}
      	\end{overpic}
\end{minipage}
\begin{minipage}[t]{0.41\linewidth}
	\begin{overpic}[width=\linewidth]{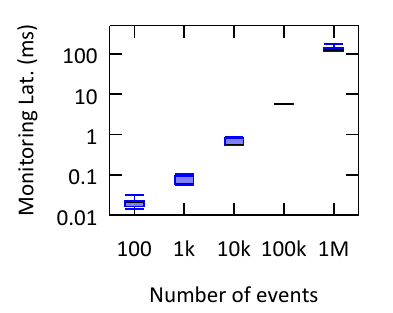}
        		\put(0,5){(b)}
      	\end{overpic}
\end{minipage}%
\vspace{-0.3cm}
\caption{Latency of (a) scheduling and (b) updating statistics.}
\label{fig:scheduling_latency}
\vspace{-0.4cm}
\end{figure}

\textbf{Scalability.}
We evaluate the scalability of our approach in two aspects. First, the \emph{scheduling latency}, i.e., the time between the detection of the start of a new window and the scheduling decision (cf. Algorithm in Figure \ref{fig:salg}). It includes predicting the negative and positive gains (cf. Algorithm in Figure \ref{fig:gainscalc}), whose complexity is determined by the granularity of the latency model, i.e., the number of bins used in the model. We measured a very low scheduling latency of in average 0.02 ms for up to 32 bins used  (cf. Figure \ref{fig:scheduling_latency}a), which is the maximal number of bins needed in any of the scenarios that we have tested (cf. Section \ref{sec:Negative and Positive Gains}). For comparison, the median scheduling latency in reactive scheduling and in Round-Robin scheduling was both 0.004 ms. Admittedly, there is a small overhead for the model-based batch scheduling controller involved compared to the simple strategies. This is not significant in most scenarios; if scheduling would be a throughput bottleneck in the splitter, the frequency of predicting negative and positive gains could be adapted (i.e., not predicting fresh gains at each single scheduling decision), trading model accuracy against throughput.

Second, we evaluate the time needed to update the latency model with new statistics from the monitoring window, i.e., the \emph{monitoring latency}. This comprises recomputing the weights, average values and standard deviations of the bins. Using 32 bins, we measured a linear growth with the number of events in the monitoring window (cf. Figure \ref{fig:scheduling_latency}b). At 1,000,000 events in a monitoring window, updating the statistics took between 100 and 200 ms, which is a reasonable time to adapt the model to changes in the workload.

\section{Related Work}
\label{sec:relatedwork}

Complex Event Processing (CEP) has evolved as the paradigm of choice to detect and integrate events in situation-aware applications \cite{Cugola:2010:TFD:1827418.1827427, Adi:2004:ASM:988145.988150, Chakravarthy1994, wu2006high}. 
In doing so, distributed CEP (DCEP) systems~\cite{Jain:2006:DIE:1142473.1142522, Schultz-Moller:2009:DCE:1619258.1619264} distribute the detection logic over a network of operators. However, individual operators can be a bottleneck and operator parallelization is needed \cite{CastroFernandez:2013:ISO:2463676.2465282, 7024105}.
Besides data parallelization, intra-operator parallelization---also known as pipelining or state-based parallelization \cite{Balkesen:2013:RRI:2488222.2488257}---has been proposed, which is limited by the functional parallelism of an operator. Besides window-based splitting as presumed in this paper, key-based splitting \cite{Hirzel:2012:PCP:2335484.2335506, CastroFernandez:2013:ISO:2463676.2465282} has been proposed, that is splitting by a key that is encoded in the events. However, this is limited to the number of different key values, e.g., different stock symbols in an algorithmic trading scenario. Moreover, not all DCEP patterns exhibit key-based data parallelism, whereas window-based data parallelism is inherent to most of them, as DCEP operators in their very nature work on windows (cf. \cite{Cugola:2010:TFD:1827418.1827427, 7024105}).

In related work, there have been addressed different problems of assigning batches of individual events to instances of stream processing operators. Das et al. \cite{Das:2014:ASP:2670979.2670995} propose a reactive controller in order to batch a minimal number of events to an operator such that the throughput is sufficiently high to process the current workload. In their processing model, operators can aggregate larger sets of events more efficiently, so that the throughput of operators grows with the batch size. A similar problem had been studied before by Carney et al. \cite{Carney:2003:OSD:1315451.1315523}. 
Micro-batching, as used, e.g., in Spark Streaming \cite{Zaharia:2012:DSE:2342763.2342773}, provides efficient failure recovery and batch-like programming paradigms by handling streaming events as a series of fixed-sized batches. Unlike in this paper, in all of these approaches, batches are composed of individual events and not of overlapping windows. 
Balkesen and Tatbul \cite{balkesen2011scalable} recognize the trade-off of communication overhead to latency in operator instances when scheduling overlapping windows. Their analytical cost model assumes fixed processing latency of an event in a window and fixed count-based or time-based window size and slide. Further, it does not consider inter-arrival times. Hence, it is not suitable for solving the batch scheduling problem in data-parallel DCEP operators.

Elasticity in data-parallel stream processing, i.e., adapting the number of operator instances to changing workloads, is a complementary problem. Existing solutions that apply latency models often base on the assumption of fair load balancing \cite{DeMatteis:2016:KCR:2851141.2851148, 7024105, lohrmann_elastic_2015}; batch scheduling defeats this assumption, deliberately inducing a controlled load imbalance. How to use the proposed latency model of the batch scheduling controller for elasticity control is an interesting research question for future work.

Other latency models for DCEP operators have been proposed. The Mace metrics from Chandramouli et al. \cite{5767926} for latency estimation in a DCEP middleware proposes an analytical model. However, it assumes the usage of their proposed scheduling algorithm---which is not a batch scheduling algorithm. In the latency model of Zeitler and Risch, a fixed processing latency of each event is assumed \cite{zeitler2011massive}; our latency model differentiates between different event types and takes into account the overlap of windows.

Batching is also applied in other fields, like graph processing \cite{Xie:2013:FIG:2556549.2556581} and column data-stores \cite{Lamb:2012:VAD:2367502.2367518}, where it is often preferable to process or store data in batches instead of handling each single tuple separately. However, typically, optimal batch sizes are predefined, e.g., by cache sizes, so that fixed batch sizes are employed.

Scheduling algorithms in non-parallel DCEP  optimize the usage of resources like CPU and memory \cite{Kencl:2008:ALS:1373990.1373994, ilprints580} without taking into account batching of overlapping data sets. 

\vspace{-0.1cm}
\section{Conclusion}
\label{sec:conclusion}

In this paper, we have tackled the problem to batch as many subsequent overlapping windows as possible to the same operator instance in data-parallel DCEP operators subject to the constraint that the operational latency in the operator instance must not exceed a given latency bound. As the batch scheduling decisions are made on open windows, a long feedback delay between the decisions and their impact on feedback parameters is induced, making reactive scheduling approaches infeasible. Instead, we have proposed a model-based controller. Evaluations show that the controller batches an optimal amount of windows even at bursty workloads. This way, the bandwidth consumption of data-parallel DCEP operators can be significantly reduced.

\bibliographystyle{ACM-Reference-Format}
\bibliography{scheduling} 


\begin{thebibliography}{00}


\ifx \showCODEN    \undefined \def \showCODEN     #1{\unskip}     \fi
\ifx \showDOI      \undefined \def \showDOI       #1{{\tt DOI:}\penalty0{#1}\ }
  \fi
\ifx \showISBNx    \undefined \def \showISBNx     #1{\unskip}     \fi
\ifx \showISBNxiii \undefined \def \showISBNxiii  #1{\unskip}     \fi
\ifx \showISSN     \undefined \def \showISSN      #1{\unskip}     \fi
\ifx \showLCCN     \undefined \def \showLCCN      #1{\unskip}     \fi
\ifx \shownote     \undefined \def \shownote      #1{#1}          \fi
\ifx \showarticletitle \undefined \def \showarticletitle #1{#1}   \fi
\ifx \showURL      \undefined \def \showURL       #1{#1}          \fi
\providecommand\bibfield[2]{#2}
\providecommand\bibinfo[2]{#2}
\providecommand\natexlab[1]{#1}
\providecommand\showeprint[2][]{arXiv:#2}

\bibitem[\protect\citeauthoryear{Adi and Etzion}{Adi and Etzion}{2004}]%
        {Adi:2004:ASM:988145.988150}
\bibfield{author}{\bibinfo{person}{Asaf Adi} {and} \bibinfo{person}{Opher
  Etzion}.} \bibinfo{year}{2004}\natexlab{}.
\newblock \showarticletitle{Amit - the Situation Manager}.
\newblock \bibinfo{journal}{{\em The VLDB Journal\/}} \bibinfo{volume}{13},
  \bibinfo{number}{2} (\bibinfo{date}{May} \bibinfo{year}{2004}),
  \bibinfo{pages}{177--203}.
\newblock
\showISSN{1066-8888}


\bibitem[\protect\citeauthoryear{Arasu, Babu, and Widom}{Arasu
  et~al\mbox{.}}{2006}]%
        {Arasu:2006:CCQ:1146461.1146463}
\bibfield{author}{\bibinfo{person}{Arvind Arasu}, \bibinfo{person}{Shivnath
  Babu}, {and} \bibinfo{person}{Jennifer Widom}.}
  \bibinfo{year}{2006}\natexlab{}.
\newblock \showarticletitle{The {CQL} Continuous Query Language: Semantic
  Foundations and Query Execution}.
\newblock \bibinfo{journal}{{\em The VLDB Journal\/}} \bibinfo{volume}{15},
  \bibinfo{number}{2} (\bibinfo{date}{June} \bibinfo{year}{2006}),
  \bibinfo{pages}{121--142}.
\newblock


\bibitem[\protect\citeauthoryear{Babcock, Babu, Datar, Motwani, and
  Thomas}{Babcock et~al\mbox{.}}{2004}]%
        {ilprints580}
\bibfield{author}{\bibinfo{person}{Brian Babcock}, \bibinfo{person}{Shivnath
  Babu}, \bibinfo{person}{Mayur Datar}, \bibinfo{person}{Rajeev Motwani}, {and}
  \bibinfo{person}{Dilys Thomas}.} \bibinfo{year}{2004}\natexlab{}.
\newblock \showarticletitle{Operator Scheduling in Data Stream Systems}.
\newblock \bibinfo{journal}{{\em The VLDB Journal\/}} \bibinfo{volume}{13},
  \bibinfo{number}{4} (\bibinfo{date}{Dec.} \bibinfo{year}{2004}),
  \bibinfo{pages}{333--353}.
\newblock
\showISSN{1066-8888}


\bibitem[\protect\citeauthoryear{Balkesen, Dindar, Wetter, and Tatbul}{Balkesen
  et~al\mbox{.}}{2013}]%
        {Balkesen:2013:RRI:2488222.2488257}
\bibfield{author}{\bibinfo{person}{Cagri Balkesen}, \bibinfo{person}{Nihal
  Dindar}, \bibinfo{person}{Matthias Wetter}, {and} \bibinfo{person}{Nesime
  Tatbul}.} \bibinfo{year}{2013}\natexlab{}.
\newblock \showarticletitle{{RIP}: Run-based intra-query parallelism for
  scalable complex event processing} {\em (\bibinfo{series}{DEBS '13})}.
  \bibinfo{publisher}{ACM}, \bibinfo{pages}{3--14}.
\newblock


\bibitem[\protect\citeauthoryear{Balkesen and Tatbul}{Balkesen and
  Tatbul}{2011}]%
        {balkesen2011scalable}
\bibfield{author}{\bibinfo{person}{Cagri Balkesen} {and}
  \bibinfo{person}{Nesime Tatbul}.} \bibinfo{year}{2011}\natexlab{}.
\newblock \showarticletitle{Scalable data partitioning techniques for parallel
  sliding window processing over data streams} {\em
  (\bibinfo{series}{International Workshop on Data Management for Sensor
  Networks (DMSN)})}.
\newblock


\bibitem[\protect\citeauthoryear{Ballani, Costa, Karagiannis, and
  Rowstron}{Ballani et~al\mbox{.}}{2011}]%
        {Ballani:2011:TPD:2018436.2018465}
\bibfield{author}{\bibinfo{person}{Hitesh Ballani}, \bibinfo{person}{Paolo
  Costa}, \bibinfo{person}{Thomas Karagiannis}, {and} \bibinfo{person}{Ant
  Rowstron}.} \bibinfo{year}{2011}\natexlab{}.
\newblock \showarticletitle{Towards Predictable Datacenter Networks} {\em
  (\bibinfo{series}{SIGCOMM '11})}. \bibinfo{pages}{242--253}.
\newblock
\showISBNx{978-1-4503-0797-0}


\bibitem[\protect\citeauthoryear{Carney, \c{C}etintemel, Rasin, Zdonik,
  Cherniack, and Stonebraker}{Carney et~al\mbox{.}}{2003}]%
        {Carney:2003:OSD:1315451.1315523}
\bibfield{author}{\bibinfo{person}{Don Carney}, \bibinfo{person}{U\u{g}ur
  \c{C}etintemel}, \bibinfo{person}{Alex Rasin}, \bibinfo{person}{Stan Zdonik},
  \bibinfo{person}{Mitch Cherniack}, {and} \bibinfo{person}{Mike Stonebraker}.}
  \bibinfo{year}{2003}\natexlab{}.
\newblock \showarticletitle{Operator Scheduling in a Data Stream Manager} {\em
  (\bibinfo{series}{VLDB '03})}. \bibinfo{publisher}{VLDB Endowment},
  \bibinfo{pages}{838--849}.
\newblock
\showISBNx{0-12-722442-4}


\bibitem[\protect\citeauthoryear{Castro~Fernandez, Migliavacca, Kalyvianaki,
  and Pietzuch}{Castro~Fernandez et~al\mbox{.}}{2013}]%
        {CastroFernandez:2013:ISO:2463676.2465282}
\bibfield{author}{\bibinfo{person}{Raul Castro~Fernandez},
  \bibinfo{person}{Matteo Migliavacca}, \bibinfo{person}{Evangelia
  Kalyvianaki}, {and} \bibinfo{person}{Peter Pietzuch}.}
  \bibinfo{year}{2013}\natexlab{}.
\newblock \showarticletitle{Integrating Scale out and Fault Tolerance in Stream
  Processing Using Operator State Management} {\em (\bibinfo{series}{SIGMOD
  '13})}. \bibinfo{publisher}{ACM}, \bibinfo{pages}{725--736}.
\newblock


\bibitem[\protect\citeauthoryear{Chakravarthy and Mishra}{Chakravarthy and
  Mishra}{1994}]%
        {Chakravarthy1994}
\bibfield{author}{\bibinfo{person}{S. Chakravarthy} {and} \bibinfo{person}{D.
  Mishra}.} \bibinfo{year}{1994}\natexlab{}.
\newblock \showarticletitle{Snoop: An expressive event specification language
  for active databases}.
\newblock \bibinfo{journal}{{\em Data Knowl. Eng.\/}} \bibinfo{volume}{14},
  \bibinfo{number}{1} (\bibinfo{year}{1994}), \bibinfo{pages}{1--26}.
\newblock


\bibitem[\protect\citeauthoryear{Chandramouli, Goldstein, Barga, Riedewald, and
  Santos}{Chandramouli et~al\mbox{.}}{2011}]%
        {5767926}
\bibfield{author}{\bibinfo{person}{B. Chandramouli}, \bibinfo{person}{J.
  Goldstein}, \bibinfo{person}{R. Barga}, \bibinfo{person}{M. Riedewald}, {and}
  \bibinfo{person}{I. Santos}.} \bibinfo{year}{2011}\natexlab{}.
\newblock \showarticletitle{Accurate latency estimation in a distributed event
  processing system} {\em (\bibinfo{series}{ICDE '11})}.
  \bibinfo{pages}{255--266}.
\newblock


\bibitem[\protect\citeauthoryear{Cugola and Margara}{Cugola and
  Margara}{2010}]%
        {Cugola:2010:TFD:1827418.1827427}
\bibfield{author}{\bibinfo{person}{Gianpaolo Cugola} {and}
  \bibinfo{person}{Alessandro Margara}.} \bibinfo{year}{2010}\natexlab{}.
\newblock \showarticletitle{TESLA: a formally defined event specification
  language} {\em (\bibinfo{series}{DEBS '10})}. \bibinfo{publisher}{ACM},
  \bibinfo{pages}{50--61}.
\newblock
\showISBNx{978-1-60558-927-5}


\bibitem[\protect\citeauthoryear{Das, Zhong, Stoica, and Shenker}{Das
  et~al\mbox{.}}{2014}]%
        {Das:2014:ASP:2670979.2670995}
\bibfield{author}{\bibinfo{person}{Tathagata Das}, \bibinfo{person}{Yuan
  Zhong}, \bibinfo{person}{Ion Stoica}, {and} \bibinfo{person}{Scott Shenker}.}
  \bibinfo{year}{2014}\natexlab{}.
\newblock \showarticletitle{Adaptive Stream Processing Using Dynamic Batch
  Sizing} {\em (\bibinfo{series}{SOCC '14})}. \bibinfo{publisher}{ACM}, Article
  \bibinfo{articleno}{16}, \bibinfo{numpages}{13}~pages.
\newblock


\bibitem[\protect\citeauthoryear{De~Matteis and Mencagli}{De~Matteis and
  Mencagli}{2016}]%
        {DeMatteis:2016:KCR:2851141.2851148}
\bibfield{author}{\bibinfo{person}{Tiziano De~Matteis} {and}
  \bibinfo{person}{Gabriele Mencagli}.} \bibinfo{year}{2016}\natexlab{}.
\newblock \showarticletitle{Keep Calm and React with Foresight: Strategies for
  Low-latency and Energy-efficient Elastic Data Stream Processing}. In
  \bibinfo{booktitle}{{\em Proceedings of the 21st ACM SIGPLAN Symposium on
  Principles and Practice of Parallel Programming}} {\em
  (\bibinfo{series}{PPoPP '16})}. \bibinfo{publisher}{ACM},
  \bibinfo{address}{New York, NY, USA}, Article \bibinfo{articleno}{13},
  \bibinfo{numpages}{12}~pages.
\newblock
\showISBNx{978-1-4503-4092-2}


\bibitem[\protect\citeauthoryear{Greenberg, Hamilton, Jain, Kandula, Kim,
  Lahiri, Maltz, Patel, and Sengupta}{Greenberg et~al\mbox{.}}{2009}]%
        {Greenberg:2009:VSF:1592568.1592576}
\bibfield{author}{\bibinfo{person}{Albert Greenberg}, \bibinfo{person}{James~R.
  Hamilton}, \bibinfo{person}{Navendu Jain}, \bibinfo{person}{Srikanth
  Kandula}, \bibinfo{person}{Changhoon Kim}, \bibinfo{person}{Parantap Lahiri},
  \bibinfo{person}{David~A. Maltz}, \bibinfo{person}{Parveen Patel}, {and}
  \bibinfo{person}{Sudipta Sengupta}.} \bibinfo{year}{2009}\natexlab{}.
\newblock \showarticletitle{VL2: A Scalable and Flexible Data Center Network}
  {\em (\bibinfo{series}{SIGCOMM '09})}. \bibinfo{pages}{51--62}.
\newblock
\showISBNx{978-1-60558-594-9}


\bibitem[\protect\citeauthoryear{Hirzel}{Hirzel}{2012}]%
        {Hirzel:2012:PCP:2335484.2335506}
\bibfield{author}{\bibinfo{person}{Martin Hirzel}.}
  \bibinfo{year}{2012}\natexlab{}.
\newblock \showarticletitle{Partition and Compose: Parallel Complex Event
  Processing} {\em (\bibinfo{series}{DEBS '12})}. \bibinfo{pages}{191--200}.
\newblock
\showISBNx{978-1-4503-1315-5}


\bibitem[\protect\citeauthoryear{Jain, Amini, Andrade, King, Park, Selo, and
  Venkatramani}{Jain et~al\mbox{.}}{2006}]%
        {Jain:2006:DIE:1142473.1142522}
\bibfield{author}{\bibinfo{person}{Navendu Jain}, \bibinfo{person}{Lisa Amini},
  \bibinfo{person}{Henrique Andrade}, \bibinfo{person}{Richard King},
  \bibinfo{person}{Yoonho Park}, \bibinfo{person}{Philippe Selo}, {and}
  \bibinfo{person}{Chitra Venkatramani}.} \bibinfo{year}{2006}\natexlab{}.
\newblock \showarticletitle{Design, Implementation, and Evaluation of the
  Linear Road Benchmark on the Stream Processing Core} {\em
  (\bibinfo{series}{SIGMOD '06})}. \bibinfo{publisher}{ACM},
  \bibinfo{pages}{431--442}.
\newblock


\bibitem[\protect\citeauthoryear{Kencl and Le~Boudec}{Kencl and
  Le~Boudec}{2008}]%
        {Kencl:2008:ALS:1373990.1373994}
\bibfield{author}{\bibinfo{person}{Lukas Kencl} {and}
  \bibinfo{person}{Jean-Yves Le~Boudec}.} \bibinfo{year}{2008}\natexlab{}.
\newblock \showarticletitle{Adaptive Load Sharing for Network Processors}.
\newblock \bibinfo{journal}{{\em IEEE/ACM Trans. Netw.\/}}
  \bibinfo{volume}{16}, \bibinfo{number}{2} (\bibinfo{date}{April}
  \bibinfo{year}{2008}), \bibinfo{pages}{293--306}.
\newblock
\showISSN{1063-6692}


\bibitem[\protect\citeauthoryear{Koliousis, Weidlich, Castro~Fernandez, Wolf,
  Costa, and Pietzuch}{Koliousis et~al\mbox{.}}{2016}]%
        {Koliousis:2016:SWH:2882903.2882906}
\bibfield{author}{\bibinfo{person}{Alexandros Koliousis},
  \bibinfo{person}{Matthias Weidlich}, \bibinfo{person}{Raul Castro~Fernandez},
  \bibinfo{person}{Alexander~L. Wolf}, \bibinfo{person}{Paolo Costa}, {and}
  \bibinfo{person}{Peter Pietzuch}.} \bibinfo{year}{2016}\natexlab{}.
\newblock \showarticletitle{SABER: Window-Based Hybrid Stream Processing for
  Heterogeneous Architectures} {\em (\bibinfo{series}{SIGMOD '16})}.
  \bibinfo{pages}{555--569}.
\newblock
\showISBNx{978-1-4503-3531-7}


\bibitem[\protect\citeauthoryear{LaCurts, Deng, Goyal, and
  Balakrishnan}{LaCurts et~al\mbox{.}}{2013}]%
        {LaCurts:2013:CNT:2504730.2504744}
\bibfield{author}{\bibinfo{person}{Katrina LaCurts}, \bibinfo{person}{Shuo
  Deng}, \bibinfo{person}{Ameesh Goyal}, {and} \bibinfo{person}{Hari
  Balakrishnan}.} \bibinfo{year}{2013}\natexlab{}.
\newblock \showarticletitle{Choreo: Network-aware Task Placement for Cloud
  Applications}. In \bibinfo{booktitle}{{\em Proceedings of the 2013 Internet
  Measurement Conference}} {\em (\bibinfo{series}{IMC '13})}.
  \bibinfo{pages}{191--204}.
\newblock
\showISBNx{978-1-4503-1953-9}


\bibitem[\protect\citeauthoryear{Lamb, Fuller, Varadarajan, Tran, Vandiver,
  Doshi, and Bear}{Lamb et~al\mbox{.}}{2012}]%
        {Lamb:2012:VAD:2367502.2367518}
\bibfield{author}{\bibinfo{person}{Andrew Lamb}, \bibinfo{person}{Matt Fuller},
  \bibinfo{person}{Ramakrishna Varadarajan}, \bibinfo{person}{Nga Tran},
  \bibinfo{person}{Ben Vandiver}, \bibinfo{person}{Lyric Doshi}, {and}
  \bibinfo{person}{Chuck Bear}.} \bibinfo{year}{2012}\natexlab{}.
\newblock \showarticletitle{The Vertica Analytic Database: C-store 7 Years
  Later}.
\newblock \bibinfo{journal}{{\em Proc. VLDB Endow.\/}} \bibinfo{volume}{5},
  \bibinfo{number}{12} (\bibinfo{date}{Aug.} \bibinfo{year}{2012}),
  \bibinfo{pages}{1790--1801}.
\newblock
\showISSN{2150-8097}


\bibitem[\protect\citeauthoryear{Li, Maier, Tufte, Papadimos, and Tucker}{Li
  et~al\mbox{.}}{2005}]%
        {Li:2005:NPN:1058150.1058158}
\bibfield{author}{\bibinfo{person}{Jin Li}, \bibinfo{person}{David Maier},
  \bibinfo{person}{Kristin Tufte}, \bibinfo{person}{Vassilis Papadimos}, {and}
  \bibinfo{person}{Peter~A. Tucker}.} \bibinfo{year}{2005}\natexlab{}.
\newblock \showarticletitle{No Pane, No Gain: Efficient Evaluation of
  Sliding-window Aggregates over Data Streams}.
\newblock \bibinfo{journal}{{\em SIGMOD Rec.\/}} \bibinfo{volume}{34},
  \bibinfo{number}{1} (\bibinfo{date}{March} \bibinfo{year}{2005}),
  \bibinfo{pages}{39--44}.
\newblock
\showISSN{0163-5808}


\bibitem[\protect\citeauthoryear{Lohrmann, Janacik, and Kao}{Lohrmann
  et~al\mbox{.}}{2015}]%
        {lohrmann_elastic_2015}
\bibfield{author}{\bibinfo{person}{Bj\"{o}rn Lohrmann}, \bibinfo{person}{Peter
  Janacik}, {and} \bibinfo{person}{Odej Kao}.} \bibinfo{year}{2015}\natexlab{}.
\newblock \showarticletitle{Elastic {Stream} {Processing} with {Latency}
  {Guarantees}}. In \bibinfo{booktitle}{{\em 2015 {IEEE} 35th {International}
  {Conference} on {Distributed} {Computing} {Systems}}}.
  \bibinfo{pages}{399--410}.
\newblock


\bibitem[\protect\citeauthoryear{Mayer, Koldehofe, and Rothermel}{Mayer
  et~al\mbox{.}}{2015}]%
        {7024105}
\bibfield{author}{\bibinfo{person}{Ruben Mayer}, \bibinfo{person}{Boris
  Koldehofe}, {and} \bibinfo{person}{Kurt Rothermel}.}
  \bibinfo{year}{2015}\natexlab{}.
\newblock \showarticletitle{Predictable Low-Latency Event Detection with
  Parallel Complex Event Processing}.
\newblock \bibinfo{journal}{{\em Internet of Things Journal, IEEE\/}}
  \bibinfo{volume}{2}, \bibinfo{number}{4} (\bibinfo{date}{Aug}
  \bibinfo{year}{2015}), \bibinfo{pages}{274--286}.
\newblock
\showISSN{2327-4662}


\bibitem[\protect\citeauthoryear{Mayer, Tariq, and Rothermel}{Mayer
  et~al\mbox{.}}{}]%
        {TR-2016-04}
\bibfield{author}{\bibinfo{person}{Ruben Mayer},
  \bibinfo{person}{Muhammad~Adnan Tariq}, {and} \bibinfo{person}{Kurt
  Rothermel}.}
\newblock \bibinfo{booktitle}{{\em {Real-time Batch Scheduling in Data-Parallel
  Complex Event Processing}}}.
\newblock \bibinfo{type}{{T}echnical {R}eport} 2016/04.
  \bibinfo{institution}{University of Stuttgart}. \bibinfo{pages}{14} pages.
\newblock


\bibitem[\protect\citeauthoryear{Schultz-M{\o}ller, Migliavacca, and
  Pietzuch}{Schultz-M{\o}ller et~al\mbox{.}}{2009}]%
        {Schultz-Moller:2009:DCE:1619258.1619264}
\bibfield{author}{\bibinfo{person}{Nicholas~Poul Schultz-M{\o}ller},
  \bibinfo{person}{Matteo Migliavacca}, {and} \bibinfo{person}{Peter
  Pietzuch}.} \bibinfo{year}{2009}\natexlab{}.
\newblock \showarticletitle{Distributed Complex Event Processing with Query
  Rewriting} {\em (\bibinfo{series}{DEBS '09})}. Article
  \bibinfo{articleno}{4}, \bibinfo{numpages}{12}~pages.
\newblock
\showISBNx{978-1-60558-665-6}


\bibitem[\protect\citeauthoryear{Wu, Diao, and Rizvi}{Wu et~al\mbox{.}}{2006}]%
        {wu2006high}
\bibfield{author}{\bibinfo{person}{Eugene Wu}, \bibinfo{person}{Yanlei Diao},
  {and} \bibinfo{person}{Shariq Rizvi}.} \bibinfo{year}{2006}\natexlab{}.
\newblock \showarticletitle{High-performance Complex Event Processing over
  Streams} {\em (\bibinfo{series}{SIGMOD '06})}. \bibinfo{pages}{407--418}.
\newblock
\showISBNx{1-59593-434-0}


\bibitem[\protect\citeauthoryear{Xie, Wang, Bindel, Demers, and Gehrke}{Xie
  et~al\mbox{.}}{2013}]%
        {Xie:2013:FIG:2556549.2556581}
\bibfield{author}{\bibinfo{person}{Wenlei Xie}, \bibinfo{person}{Guozhang
  Wang}, \bibinfo{person}{David Bindel}, \bibinfo{person}{Alan Demers}, {and}
  \bibinfo{person}{Johannes Gehrke}.} \bibinfo{year}{2013}\natexlab{}.
\newblock \showarticletitle{Fast Iterative Graph Computation with Block
  Updates}.
\newblock \bibinfo{journal}{{\em Proc. VLDB Endow.\/}} \bibinfo{volume}{6},
  \bibinfo{number}{14} (\bibinfo{date}{Sept.} \bibinfo{year}{2013}),
  \bibinfo{pages}{2014--2025}.
\newblock
\showISSN{2150-8097}


\bibitem[\protect\citeauthoryear{Zaharia, Das, Li, Shenker, and Stoica}{Zaharia
  et~al\mbox{.}}{2012}]%
        {Zaharia:2012:DSE:2342763.2342773}
\bibfield{author}{\bibinfo{person}{Matei Zaharia}, \bibinfo{person}{Tathagata
  Das}, \bibinfo{person}{Haoyuan Li}, \bibinfo{person}{Scott Shenker}, {and}
  \bibinfo{person}{Ion Stoica}.} \bibinfo{year}{2012}\natexlab{}.
\newblock \showarticletitle{Discretized Streams: An Efficient and
  Fault-tolerant Model for Stream Processing on Large Clusters} {\em
  (\bibinfo{series}{HotCloud'12})}. \bibinfo{publisher}{USENIX Association}.
\newblock


\bibitem[\protect\citeauthoryear{Zeitler and Risch}{Zeitler and Risch}{2011}]%
        {zeitler2011massive}
\bibfield{author}{\bibinfo{person}{Erik Zeitler} {and} \bibinfo{person}{Tore
  Risch}.} \bibinfo{year}{2011}\natexlab{}.
\newblock \showarticletitle{Massive scale-out of expensive continuous queries}.
\newblock \bibinfo{journal}{{\em VLDB Endowment\/}} \bibinfo{volume}{4},
  \bibinfo{number}{11} (\bibinfo{year}{2011}), \bibinfo{pages}{1181--1188}.
\newblock


\end{thebibliography}

\end{document}